\documentclass{emulateapj} 
\usepackage{apjfonts} 
\usepackage{html}
\newcommand{\smjustify}{\hspace{-0.1cm}}
\def\smwidth{0.5\textwidth}
\def\smhwidth{0.5\textwidth}
\submitted{Accepted for Publication in ApJ}

\def\gsim{\mathrel{\rlap{\lower2pt\hbox{\hskip0pt\small$\sim$}}
\raise2pt\hbox{\small $>$}}}
\def\lsim{\mathrel{\rlap{\lower2pt\hbox{\hskip0pt\small$\sim$}}
\raise2pt\hbox{\small $<$}}}     \def\bq{\begin{equation}}
\def\eq{\end{equation}}

\shorttitle{The Opacity of the IGM} \shortauthors{EMBERSON, THOMAS \& ALVAREZ}

\begin{document}

\title{The Opacity of the Intergalactic Medium during Reionization: Resolving Small-Scale Structure}    

\author{J.D.~Emberson\altaffilmark{1,2}, Rajat
M. Thomas\altaffilmark{1} and Marcelo A. Alvarez\altaffilmark{1}}

\altaffiltext{1}{Canadian Institute for Theoretical Astrophysics,
University of Toronto, 60 St. George St., Toronto, ON M5S 3H8, Canada}
\altaffiltext{2}{Department of Astronomy and Astrophysics, University
of Toronto, 50 St. George, Toronto, ON M5S 3H4, Canada}
\email{emberson@astro.utoronto.ca}

\begin{abstract}

Early in the reionization process, the intergalactic medium (IGM)
would have been quite inhomogeneous on small scales, due to the low
Jeans mass in the neutral IGM and the hierarchical growth of structure
in a cold dark matter Universe. This small-scale structure acted as an
important sink during the epoch of reionization, impeding the progress
of the ionization fronts that swept out from the first sources of
ionizing radiation. Here we present results of high-resolution
cosmological hydrodynamics simulations that resolve the cosmological
Jeans mass of the neutral IGM in representative volumes several Mpc
across. The adiabatic hydrodynamics we follow are appropriate in an
{\em unheated} IGM, before the gas has had a chance to respond to the
photoionization heating. Our focus is determination of the resolution
required in cosmological simulations in order to sufficiently sample
and resolve small-scale structure regulating the opacity of an
unheated IGM. We find that a dark matter particle mass of $m_{\rm dm}
\lesssim 50~M_\odot$ and box size of $L \gtrsim 1$ Mpc are
required. With our converged results we show how the mean free path of
ionizing radiation and clumping factor of ionized hydrogen depends
upon the ultraviolet background (UVB) flux and redshift. We find, for
example at $z=10$, clumping factors typically of 10 to 20 for an
ionization rate of $\Gamma \sim 0.3 - 3 \times 10^{-12}$s$^{-1}$, with
corresponding mean free paths of $\sim 3-15$~Mpc, extending previous
work on the evolving mean free path to considerably smaller scales and
earlier times.

\end{abstract}


\section{Introduction}

The fact that the most abundant sources of radiation during reionization 
are likely to be currently undetectable \citep[e.g.,][]{trenti/etal:2010, oesch/etal:2012,
alvarez/etal:2012} means that the details of the reionization process are beyond 
most current observational probes.
The notable exceptions are observations of the polarization of
the cosmic microwave background (CMB), which imply an optical depth to
Thomson scattering of $\tau\sim 0.09$ \citep{komatsu/etal:2011}, and
the appearance of a Gunn-Peterson trough in the spectra of distant
quasars \citep{Fan06}, indicating that reionization was largely
complete by $z\sim 6$. Reionization is therefore thought to have
mainly taken place over the redshift range $z\sim 6-15$. Due to the
lack of more specific constraints, much of our current understanding
about the epoch of reionization comes from theoretical studies in the
context of the $\Lambda$CDM cosmology. 

The picture which always emerges is of small-scale gaseous structures forming
at $z>20$, due to the collapse of dark matter halos at the Jeans
scale, roughly $10^4~M_\odot$
\citep[e.g.,][]{peebles/dicke:1968,couchman/rees:1986,shapiro/etal:1994,gnedin/hui:1998}. The
gas was just cool enough to fall into halos at this mass, leading to
strong inhomogenities on a scale of tens of comoving parsecs. At the
same time, slightly more massive halos, with masses on the order of
$\sim 10^6~M_\odot$, formed enough H$_2$ molecules in their cores to
cool efficiently, leading to the formation of the first stars in the
Universe
\citep[e.g.,][]{tegmark/etal:1997,2002Sci...295...93A,2002ApJ...564...23B,2003ApJ...592..645Y}.
The ionizing radiation from these stars is thought to have created
substantial, yet short-lived H~II regions, which were shaped by the
surrounding inhomogeneity of the  gas distribution
\citep{alvarez/etal:2006,abel/etal:2007,yoshida/etal:2007}. 

Eventually, sufficiently large halos formed that triggered the
formation of the first galaxies
\citep{johnson/etal:2007,wise/abel:2008,greif/etal:2008}. These
nascent dwarf galaxies would have created longer-lived and isolated
H~II regions \citep{wise/cen:2009,wise/etal:2012}. It is unclear how
these galaxies evolved into the much more luminous ones that have been
observed at redshifts as high as $z\sim 8$
\citep[e.g.,][]{bouwens/etal:2010}.  Nevertheless, it is widely
believed that as the first galaxies grew and merged, their collective
radiative output created a large and complex patchwork of ionized
bubbles, with characteristic sizes on the order of tens to hundreds of
comoving Mpc
\citep[e.g.,][]{barkana/loeb:2004,furlanetto/etal:2004,iliev/etal:2006}. During
this time, dense systems in the IGM likely impeded the progress of
ionization fronts
\citep{barkana/loeb:1999,haiman/etal:2001,shapiro/etal:2004,iliev/etal:2005,ciardi/etal:2006}.
At the end of reionization the so-called ``Lyman-limit'' systems,
dense clouds of gas optically-thick to ionizing radiation observed in
the spectra of quasars at $z<6$
\citep[e.g.,][]{storrie-lombardi/etal:1994,prochaska/etal:2009},
dominated the overall opacity of the IGM to ionizing radiation. These
systems crucially influenced the percolation phase of reionization
\citep{gnedin/fan:2006,choudhury/etal:2009,alvarez/abel:2012}, which
in turn determined the evolution and structure of the ionizing
background
\citep[e.g.,][]{haardt/madau:1996,bolton/haehnelt:2007a,mcquinn/etal:2011}.

Thus, the progress of reionization depended not only on the properties
of the sources of ionizing radiation, but also on the {\em
  sinks}. Theoretical models of reionization must describe not just
the spectral energy distribution, abundance, and clustering of early
sources of ionizing radiation, but also the inhomogeneity of the
intergalactic medium (IGM) in the space between the sources. It is
this latter description that is the goal of the present work.

Early descriptions of reionization took into account inhomogeneities
in the IGM through a ``clumping factor'', $c_l$, by which the
recombination rate is boosted relative to the homogeneous case. This
allows one to write a global ionization rate, equal to the ionizing
photon emmissivity minus the recombination rate of a clumpy IGM, and
thereby determine the reionization history for a given ionizing source
population. \citet{shapiro/giroux:1987} used such a model to show that
the observed population of QSOs were insufficient to have reionized
the Universe by $z\sim 5$. Their assumption of  $c_l\sim 1$ would
have been conservative, in that that additional recombinations would
have made it even more difficult for quasars alone to reionize the
Universe.

In addition to being useful in modeling the reionization history, the
clumping factor is also important in estimating the necessary number
of ionizing photons per baryon to {\em maintain} an ionized
Universe. The necessary and sufficient condition for maintaining an
ionized Universe is that the ionizing photon emissivity should be
greater than or equal to the recombination rate of the
IGM. \citet{madau/etal:1999} used this fact to derive a critical star
formation rate, above which the rate of ionizing photons is enough to
maintain the Universe in an ionized state.

\citet{gnedin/ostriker:1997} used hydrodynamic simulations with
a treatment of photoionization in the ``local optical depth''
approximation to determine the clumping factor of the ionized
component of the IGM, finding a value of $c_l\sim 30$ at $z=6$. They
also pointed out that the actual clumping factor of the IGM
would have been larger due to structure on smaller scales than they
resolved. 
More recently, \citet{miralda-escude/etal:2000} built a
semi-analytical model for the reionization of an inhomogenous IGM, in
which the underlying gas density distribution was determined by
numerical simulations.  They argued that in addition to specifying the
clumping factor of the ionized medium, it is also necessary to
describe the distribution of high-density gas clouds that are able to
self-shield against ionizing radiation.

\cite{mcquinn/etal:2011} followed a similar approach to that of
\cite{miralda-escude/etal:2000} to  explain the evolution of the
ionizing background radiation at redshifts less than $z\sim 6$, using
more realistic numerical simulations which were post-processed with
radiative transfer. These works were focused on the large scales
relevant in the post-reionization IGM, after photoionization heating
has ``ironed out'' the clumpiness of the IGM on the smallest
scales. The timescale over which this smoothing occurs is on the order
of $10-100$ Myr \citep{2005MNRAS.361..405I}. Although the recombination 
rate in the homogenous IGM is on the order of 1 Gyr, 
small-scale inhomogeneities increase recombinations
by at least an order of magnitude, making the recombination time 
of the high-redshift IGM comparable to the smoothing time.
Our work here is focused on the higher redshifts and smaller
scales that were most relevant early in reionization before much smoothing
has occurred.

We seek to obtain convergence in the quantities that describe the inhomogeneity of the {\em
  unheated} IGM during the epoch of reionization, such as the mean free
path, $\lambda$, clumping factor, $c_l$, and density threshold above
which gas is self-shielded, $\Delta_{\rm crit}$, by spanning the
parametre space of redshift and ionizing background intensity,
$j_\nu$. To do this, we post-process cosmological adiabatic
hydrodynamics simulations with radiative transfer calculations along
different lines of sight through the simulated volume. Radiative
feedback raises the Jeans mass of the IGM, thereby increasing the
scale of inhomogeneities. Therefore, the resolutions we find in our
adiabatic simulations that are necessary to resolve structure in the
unheated IGM are also sufficient to model radiative feedback at
all times.

The outline of the paper is as follows. Details of the simulation
setup and radiative transfer are described in $\S 2$. In $\S 3$ we
present our numerical results, followed by $\S 4$, where we present
the results of our convergence tests. $\S 5$ concludes with a
discussion of our main results.

\begin{table}[t] 
\caption{Simulation Parameters}
\centering
\begin{tabular}{c c c c c c c c}
\hline\hline
Simulation & $N$ & $L$ (Mpc) & $m_{\rm dm}$ ($M_\odot$) & $r_{\rm soft}$ (pc) \\
\hline
A1 & $2\times 256^3$ & 0.25 & 31                 & 30  \\
A3 & $2\times 256^3$ & 1    & $2.0 \times 10^3$  & 120 \\
A4 & $2\times 256^3$ & 2    & $1.6 \times 10^4$  & 240 \\
A6 & $2\times 256^3$ & 8    & $1.0 \times 10^6$  & 960 \\
\hline
B1 & $2\times 512^3$ & 0.25 & 3.8                & 15  \\
B2 & $2\times 512^3$ & 0.5  & 31                 & 30  \\
B3 & $2\times 512^3$ & 1    & 240                & 60  \\
B4 & $2\times 512^3$ & 2    & $2.0 \times 10^3$  & 120 \\
B5 & $2\times 512^3$ & 4    & $1.6 \times 10^4$  & 240 \\
B6 & $2\times 512^3$ & 8    & $1.3 \times 10^5$  & 480 \\
\hline
C1 & $2\times 1024^3$ & 0.25 & 0.5               & 7.5 \\
C2 & $2\times 1024^3$ & 0.5  & 3.8             & 15 \\
C3 & $2\times 1024^3$ & 1    & 31                & 30  \\
C4 & $2\times 1024^3$ & 2    & 240               & 60  \\
C6 & $2\times 1024^3$ & 8    & $1.6 \times 10^4$ & 240 \\
\label{table:simparams}
\end{tabular}
\end{table}
   
\section{Numerical Approach}

Here we describe our numerical approach, in which we perform a suite of
cosmological adiabatic\footnote{In this case the gas cools adiabatically 
with the expansion of space while radiative heating and cooling 
processes that would otherwise affect its temperature are ignored.
The justification and consequences for assuming this choice are discussed in the text.} 
hydrodynamics simulations using the publicly 
available SPH code Gadget-2 \citep{springel/etal:2005}. 
We then postprocess each simulation with multifrequency 
radiative transfer of hydrogen ionizing radiation, assuming
photoionization equilibrium, to determine the dependence of basic
quantities, like the ionizing photon mean free path and clumping
factor, on redshift and intensity of the background radiation field.
The mean free path presented here is used to quantify the opacity
of the IGM to ionizing radiation and should be considered a local
quantity that depends on the spatial variation of the UVB flux 
during patchy reionization.

\subsection{Cosmological Hydrodynamic Simulations}

The cosmological simulations are parameterized by box size, $L$,
and total number of dark matter and gas particles, $N$. Table
\ref{table:simparams} summarizes these parameters for our suite of
simulations and lists their corresponding dark matter particle masses,
$m_{\rm dm}$, along with the comoving gravitational softening length,
$r_{\rm soft}$. The simulations were evolved from redshift $z = 200$
to $z = 6$, except for simulations C1 through C4 which, due to computational
limitations, were terminated early at $z = 10$. Initial conditions
were generated separately for dark matter and baryons using
transfer functions computed by CAMB for each component, with the same
random phases. Throughout our work we assume the set of cosmological
parameters ($\Omega_{\mathrm{DM}}$, $\Omega_\mathrm{b}$,
$\Omega_\Lambda$, $h$) = (0.228, 0.042, 0.73, 0.72).  

\begin{figure} \smjustify
\includegraphics[width=\smwidth]{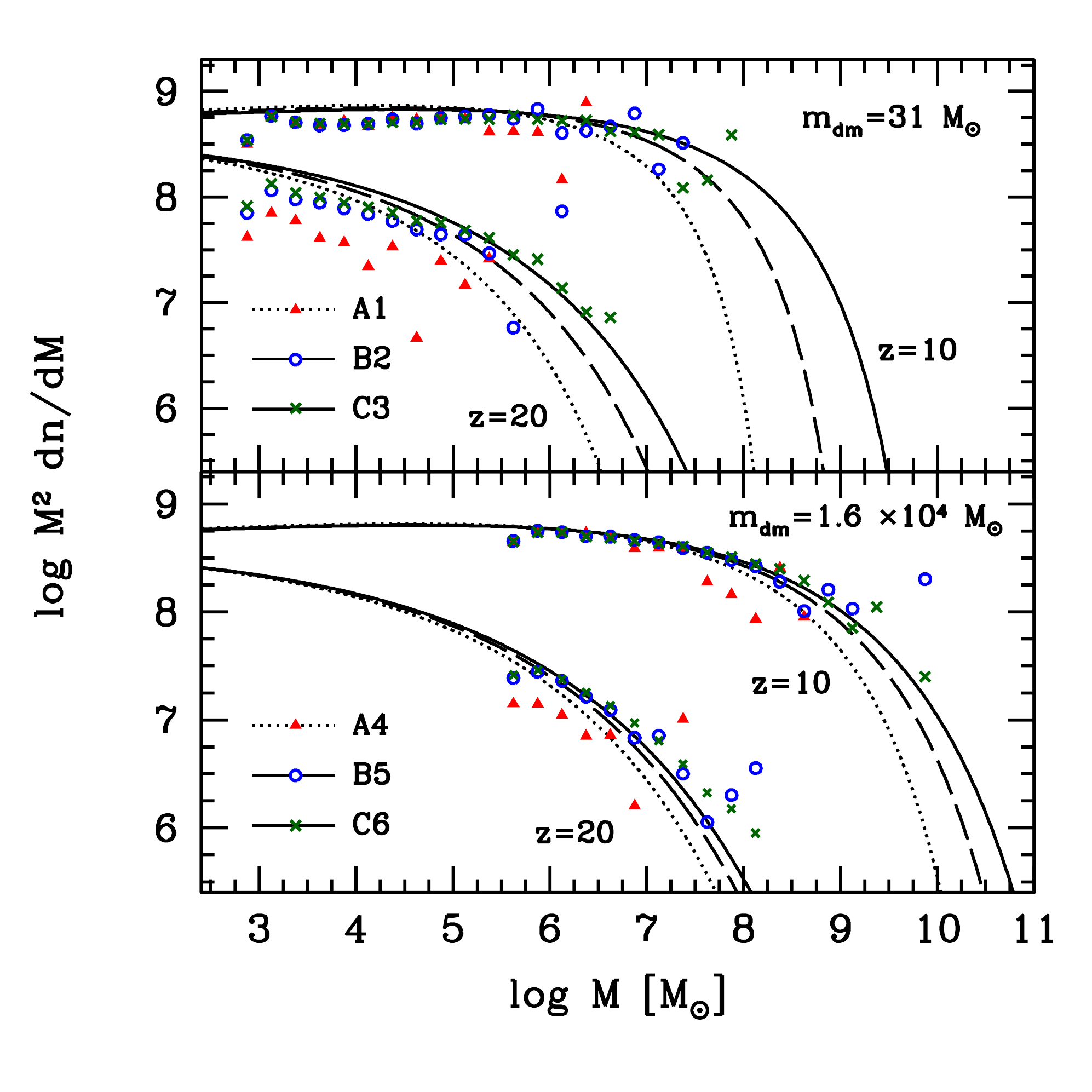} \vspace{-0.7cm}
\caption{Halo mass functions from six simulations are compared to
their Warren et al. counterparts using the variance
$\sigma_{\rm eff}^2 \equiv \sigma^2 - \sigma^2(M_{\rm box})$
at redshifts 10 and 20.  The top
panel plots simulations A1, B2, and C3 each containing dark matter
masses of $m_{\rm dm} = 31~M_\odot$ while the lower panel plots
simulations A4, B5, and C6 with $m_{\rm dm} = 1.6 \times 10^4~
M_\odot$. In each case, points denote halo mass functions obtained
from the simulations while the lines trace the corresponding Warren et al.
curves.}
\vspace{0.2cm}   
\label{figure:hmfs}
\end{figure}

A quantitative test of the simulated structure formation is to
identify dark matter halos to construct mass
functions, $dn/dM$, which can be compared to analytic models. 
Figure \ref{figure:hmfs} shows the mass
functions obtained from a friends-of-friends (FOF) halo
identification scheme with linking length of 0.2 mean interparticle
spacings, at redshifts $z = 10$ and 20 for two groups of
simulations sharing common mass resolutions of $m_{\rm dm} = 31~
M_\odot$ and $1.6 \times 10^4~M_\odot$. A common fitting function to
compare to is the \citet{warren/etal:2006} mass function. When doing
so, however, it is important to note that this model assumes a Universe with
infinite spatial extent; something that cannot be achieved using
numerical simulations. It is therefore useful to compute Warren et al.
mass functions using a modified variance of the form  
$\sigma_{\rm eff}^2 \equiv \sigma^2 - \sigma^2(M_{\rm box})$, 
where  $M_{\rm box}$ is the total mass
contained within the simulated volume. This has the effect of removing
contributions from mass fluctuations on scales larger than
that of the simulated volume. With this correction we find that the
Warren et al. mass function is generally well-matched by the
numerical simulations. 

There is one important feature worth noting in Figure
\ref{figure:hmfs}: For fixed mass resolution, simulations with 
larger volumes tend to trace the
analytic curves more closely. This is most noticeable in the top panel
for $z = 20$. We can attribute this to the fact that at fixed
resolution, simulations  with larger volumes will contain a more
statistically representative collection of halos.  In $\S$4 we will
show how sample variance in small boxes has important consequences for
numerical convergence. Even though a simulation may have a sufficient
mass resolution to resolve low-mass halos within the IGM, its volume
may be so small that sample variance causes noticeable variation
in computed quantities between different random realizations.
Recall that we used a modified variance $\sigma_{\rm eff}^2$
when computing the analytic curves of Figure \ref{figure:hmfs} in
order to compensate for missing large-scale power in the finite volume boxes.
Though not necessary for our purposes here, we point out to the reader that
\citet{reed/etal:2007} present an additional correction
that can be used to adjust simulated mass functions for sample variance 
from small volumes.

It is well known that the inhomogeneous nature of the IGM plays an
important role in the progression of the reionization epoch. This was
emphasized by \citet{miralda-escude/etal:2000} who presented an
evolutionary model of reionization based on the gas density
distribution observed in numerical simulations. It is therefore
useful to examine the density distribution of baryons within
cosmological simulations, through the use of the probability density
function (PDF), $P(\Delta)$, defined to be the normalized distribution
of gas in terms of overdensity $\Delta\equiv \rho/{\overline{\rho}}$. 

\begin{figure}
\smjustify
\includegraphics[width=\smwidth]{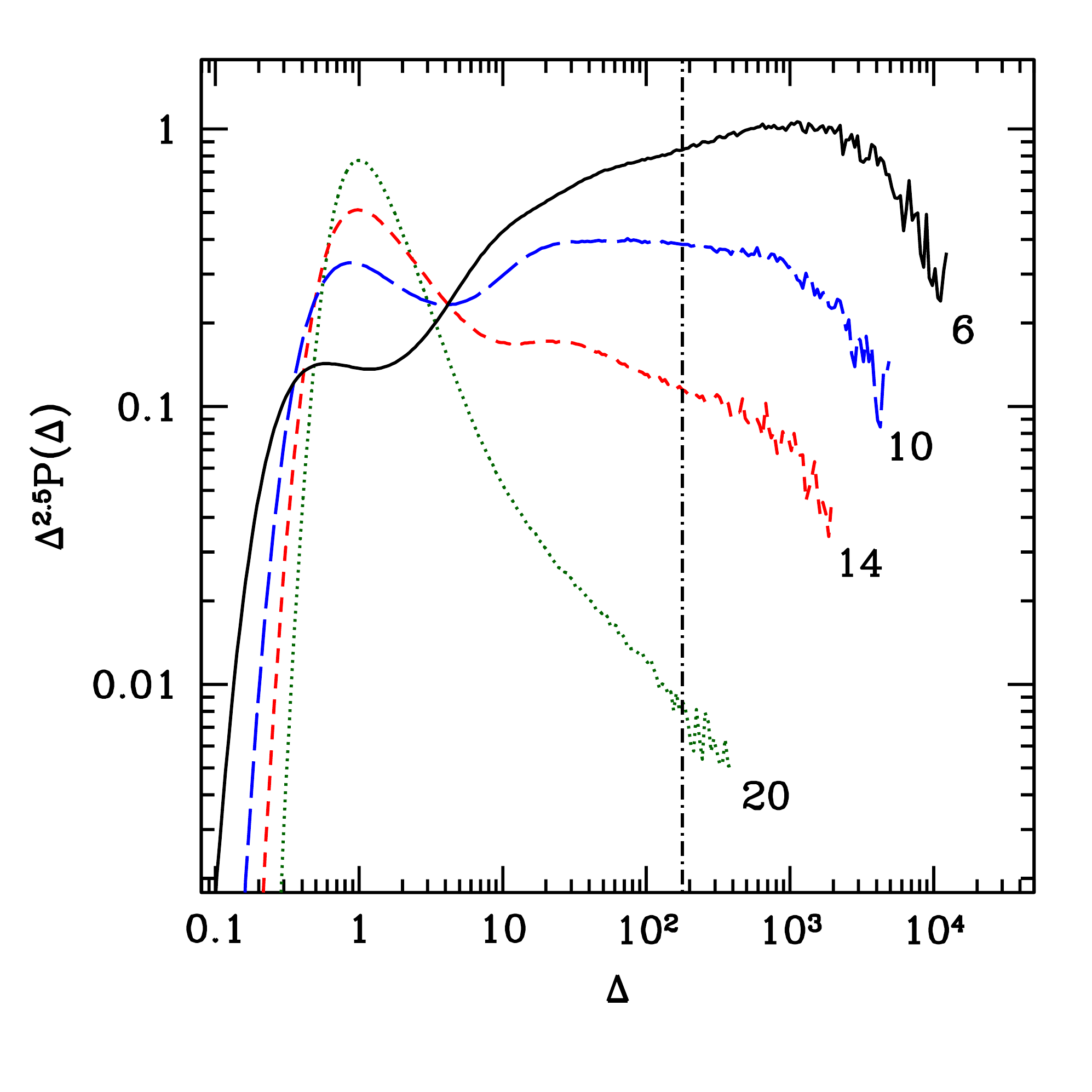}
\vspace{-1cm}
\caption{Gas PDFs for our fiducial simulation B2 ($512^3$, 0.5 Mpc)
at different redshifts, as labelled beside the curves. The vertical
dot-dashed line delineates the overdensity $\Delta_{\rm vir} = 18 \pi^2$ 
of a virialized isothermal sphere. If collapsed structures have a density 
profile of the form $\rho\propto r^{-2}$,
for $\Delta_{\rm vir} < \rho / \bar{\rho} < \Delta_{\rm max}$, then 
$P(\Delta)\propto \Delta^{-2.5}$ for $\Delta_{\rm vir} < 
\Delta < \Delta_{\rm max}$. We plot the PDF multiplied by 
$\Delta^{2.5}$ so that the curve should approach a constant for 
$\Delta > \Delta_{\rm vir}$ if gas is collapsed within these structures.}
\label{figure:pdfs}
\vspace{0.2cm}  
\end{figure}

In Figure \ref{figure:pdfs} we plot volume-weighted gas PDFs from our
fiducial simulation B2 which contains $2 \times 512^3$ dark matter
plus gas particles in a box of comoving length 0.5 Mpc. More
precisely, we plot $\Delta^{2.5}P(\Delta)$ which is expected  to
approach a constant at $\Delta > \Delta_{\rm vir} = 18 \pi^2$ if gas
at those densities is collapsed  within halos described by a density
profile $\rho\propto r^{-2}$. As time evolves, the fraction of gas
collapsed within halos increases, though only for $z \lesssim 10$ does
the high-density tail of the PDF appear to approach a constant value.

Recall that the PDFs shown here correspond to adiabatic simulations
for which radiative heating and cooling processes are ignored. 
We expect that heating would evolve the gas distribution in such a way as to 
decrease the amplitude of $P(\Delta)$ at large values of $\Delta$ as gas
boils out of overdense regions. In this work we are concerned with the
initial phase of reionization before substantial heating occurs, and
discuss at length in $\S5$ how heating would affect our results. In contrast,
radiative cooling would act to promote $P(\Delta)$ at large 
values of $\Delta$ from the enhanced collapse of gas into overdense halos. 
However, our work is primarily concerned with determining the resolution
requirements for minihalos  with masses on the order of $10^4~M_\odot$. For these halos,
H$_2$ cooling is the dominant mechanism, but this is rapidly suppressed
once an UVB is established within the IGM \citep{haiman/etal:2000}. Radiative
cooling is thus most relevant for the easily resolved large halos, and our choice to omit
cooling should not qualitatively affect our conclusions with respect to
convergence of small-scale structure.

\subsection{Post-processed Ionization Calculation}

In order to simulate the effects of self-shielding by absorption
systems, we postprocess
the SPH density field with a multifrequency radiative transfer algorithm. This
involves tracing  the attenuation of the ionizing radiation along
different lines of sight  throughout the volume while assuming
photoionization equilibrium. 

\subsubsection{Ultraviolet Background Spectrum}

We consider a background ionizing intensity $I_\nu$, 
so that the flux of photons capable of ionizing hydrogen is
\bq
F = \int d\Omega\int_{\nu_{HI}}^{4\nu_{HI}} \frac{I_\nu}{h\nu} d\nu,
\label{flux}
\eq
where $h\nu_{HI}=13.6$~eV is the photon energy at the Lyman
edge. The upper limit in the integral corresponds to the ionizing
threshold for fully ionizing helium -- we assume helium is singly
ionized along with hydrogen, and therefore only consider photons below
the He~II Lyman edge. 

We adopt a power-law UVB spectrum,
\bq
I_\nu = I_o \left( \frac{\nu}{\nu_{HI}}  \right)^{-\alpha},
\label{eq:intensitybox} 
\eq 
where $\nu_{HI} \leq \nu \leq 4\nu_{HI}$
and $I_o$ is the intensity at the Lyman edge. In our analysis we have
sampled a region of parameter space for which $1 \leq \alpha \leq
3$. Our fiducial value of $\alpha = 2$ is chosen to be consistent with
the spectral index we would  expect for an ionizing background
produced from a mixture of galaxies and quasars
\citep[e.g.,][]{bolton/haehnelt:2007a}. Our results exhibit only a minor
dependence on spectral index in this range, as also found by
\citet{mcquinn/etal:2011}. For this reason we henceforth refer only to
our fiducial case of $\alpha = 2$.  

The intensity is often expressed in terms of the quantity $J_{-21}$,
defined to be the isotropic equivalent of $I_o$, $(\int
I_0d\Omega)/(4\pi)$, in units of $10^{-21}\ {\rm erg}\ {\rm
cm}^{-2}{\rm  s}^{-1} {\rm Hz}^{-1}{\rm ster}^{-1}$. For the form
expressed in equation (\ref{eq:intensitybox}), we can integrate
equation (\ref{flux}) to relate $J_{-21}$ to the flux of ionizing
photons. For $\alpha=2$, we obtain: \bq J_{-21} =  0.09 \left(
\frac{F}{10^5\ {\rm cm}^{-2}{\rm s}^{-1}} \right).
\label{eq:J21} \eq Another useful quantity to describe the UVB is
$\Gamma_{-12}$, defined to be the ionization rate per atom, in units
of $10^{-12}\ {\rm s}^{-1}$:
\bq \Gamma_{-12} = 0.3 \left(
\frac{F}{10^5\ {\rm cm}^{-2}{\rm s}^{-1}} \right).
\label{eq:gamma12}
\eq
Note that this refers to the ionization rate corresponding to a given
background, and not the mean ionization rate per atom along our rays, 
which is lower due to attenuation and includes the neutral component
of the IGM. 

We calculate the ionization state of the volume for a broad range of
background 
flux with the fiducial value of $F = 10^{5}\ {\rm cm}^{-2} {\rm s}^{-1}$ 
taken to be consistent with the value of $\Gamma_{-12} \sim 0.3$ inferred
from the optical depth of the Ly$\alpha$ forest seen in quasar spectra
\citep[e.g.,][]{bolton/haehnelt:2007a}. Due to its common usage in the
literature, we will report our results in terms of $\Gamma_{-12}$,
though it should be remembered that its conversion to flux simply
follows equation (\ref{eq:gamma12}). 

\subsubsection{Ray-tracing}

In our ray-based approach, the UVB has a plane-parallel
direction dependence, so that $I_\nu = F_\nu\delta(\hat{n})$, where
$\hat{n}$ is the direction of propagation of the radiation and $F_\nu$
is the spectral flux density. This is appropriate especially in the
beginning stages of reionization, where a given patch of the IGM is
initially exposed to a one-sided flux from the downstream direction of
the ionization front. 
In addition, we use the ``case B'' recombination coefficient which
assumes that recombinations to the ground state are quickly canceled by
subsequent photoionizations and implies that rays can be treated 
independently. In reality, ionizing radiation produced by recombinations 
directly to the ground state becomes part of the UVB, changing the spectral 
shape that we adopt in equation (\ref{eq:intensitybox}), but not the total ionizing 
photon flux. Given the relative insensitivity we find to the spectral shape, 
using case B recombination rates should be a good approximation.
Finally, because the equilibration time is very short
compared to the Hubble time, we use photoionization equilibrium, which
allows us to calculate the ionization state and attenuation of the
background self-consistently by sequentially iterating along the ray
in the direction $\hat{n}$. 

To obtain an unbiased sample of the gas density field and minimize
noise, the rays are assigned starting points uniformly distributed in a
plane with orientations perpendicular to the plane. We use three
orthogonal planes in order to sample different directions.
Each ray segment corresponds to a cubic volume element, within which
the mean density is obtained from the SPH particle data by the
mass-conserving spline interpolation outlined in
\citet{alvarez/etal:2006}. The ray segments have lengths given by
$L/N_{\rm ray}$, where $L$ is the box size, so that the number of rays is
proportional to $N_{\rm ray}^2$, while the number of segments along a
given ray is proportional to $N_{\rm ray}$. We check for convergence 
in our radiative transfer calculations by interpolating to a variety of 
values for $N_{\rm ray}$. From this we find that it is necessary to interpolate
to $N_{\rm ray} = 1024$ for the $256^3$ and $512^3$ particle simulations
and to $N_{\rm ray} = 2048$ for the $1024^3$ particle simulations.

\subsubsection{Equilibirum Radiative Transfer}

The equation of radiative transfer for $I_\nu$ is
\bq
\frac{dI_\nu}{ds}=-n_{\rm HI}{\sigma_\nu}I_\nu+\epsilon_\nu,
\label{transfer}
\eq
where $s$ is the proper distance and $n_{\rm HI}$ is the proper number
density of neutral hydrogen. Here we are concerned with the transfer
of ionizing radiation through a patch of IGM in which there are no
sources, and therefore set $\epsilon_\nu=0$. 

The ionization rate within a given ray segment is related to the background
intensity and neutral hydrogen column density, $N_{HI}$, integrated
along a given ray by using the solution of equation (\ref{transfer}): 
\bq
\Gamma = 4 \pi \int_{\nu_{HI}}^{4\nu_{HI}} \frac{I_\nu}{h \nu} 
\sigma_\nu e^{-N_{HI} \sigma_\nu} d\nu, 
\label{gamma} 
\eq 
where $\sigma_\nu$ is the absorption cross section, with mean value
\bq
\bar{\sigma}\equiv\frac{\int_{\nu_{HI}}^{4\nu_{HI}}
  \sigma_\nu\frac{I_\nu}{h\nu}d\nu}   
{\int_{\nu_{HI}}^{4\nu_{HI}}\frac{I_\nu}{h\nu}d\nu} = 2.84 \times
10^{-18}~{\rm cm^2}. 
\label{eq:sigmabar}
\eq
The flux of ionizing photons is diminished along the ray through
absorption by intervening neutral hydrogen, and the spectrum steepens
as softer photons are preferentially absorbed.

To determine the opacity along the ray self-consistently, we iterate
along the ray, using the total H I column density from the previous
ray segments to calculate the photoionization rate at the current
segment using equation (\ref{gamma}). This is then used to determine
the neutral hydrogen density in the current segment under the assumption of
photoionization equilibrium:
\bq 
\Gamma n_{HI} = \alpha_B n_{HII}n_e,
\label{eq:gameqbm}
\eq  
where $n_{HI}$, $n_{HII}$ and $n_e$ are the number densities of
neutral hydrogen, ionized hydrogen, and electrons within that segment.
The resulting value of H I density is then used to
update the total column density, and the procedure is repeated until
the end of the ray is reached.

Equation (\ref{eq:gameqbm}) assumes a uniform radiation field within
each ray segment.  This assumption breaks down if the segment becomes
sufficiently optically thick that $\Gamma$ changes significantly
across it. To address this issue, individual ray segments are split
into plane-parallel subsegments in the direction $\hat{n}$ with widths
chosen such  that the flux passing through each subsegment is
attenuated by no more than $2\%$ of its initial value. Photoionization
equilibrium is applied in sequence to each subsegment and global
quantities pertaining to the segment as a whole are computed  as
volume averages over each subsegment. 

We assume that all free electrons within the volume come from hydrogen 
and consider a uniform gas temperature of $T_{\rm gas} = 10^4$ K 
so that $\alpha_B = 2.6 \times 10^{-13}\ {\rm cm}^3 {\rm s}^{-1}$.  
Including helium in our calculations would lead to small
corrections in the hardening of the radiation at high optical depths,
due to the slightly different frequency dependence of the He I
absorption cross-section relative to that of H I. Given the
insensitivity of our results to varying the spectral slope $\alpha$,
inclusion of helium radiative transfer would not improve the accuracy
of our results, while needlessly complicating their interpretation, so
we neglect it.

\subsubsection{Optimal Ray Length and the Mean Free Path}

The opacity of the IGM can be written in terms of the mean free path,
with the equation of transfer for the flux of ionizing photons given by 
\bq
\frac{dF}{ds}= - \frac{F(1+z)}{\lambda},
\label{ftransfer}
\eq
where $F$ is the total flux of ionizing photons in units of
cm$^{-2}$s$^{-1}$, and the factor $1+z$ accounts for the fact that we
define the mean free path to be in comoving units, and is the
definition we use throughout this paper. We calculate the mean free
path along a given ray as the solution to equation (\ref{ftransfer}):
\bq
\lambda = - \frac{s}{{\rm ln}(F_{\rm out} / F)},
\label{eq:slabmfp}
\eq
where $s$ is now the comoving length of the ray, $F$ is the incident
flux at the start of  the ray, and $F_{\rm out}$ is the attenuated
flux leaving the last ray segment. The total mean  free path is determined by
first averaging $F_{\rm out}$ over all rays, then applying equation
(\ref{eq:slabmfp}).   

Naively, we may choose to set $s = L$ so that each ray samples the
entire length of the box.  However, we must be careful since the use
of equation (\ref{eq:slabmfp}) is not physically meaningful in the
optically thick limit where $F_{\rm out}$ can tend to 0. In other
words, we want to determine the opacity of the IGM due to small-scale
structure at a fixed background flux, but including the cumulative
effect of this shielding over distances approaching the mean free path
would correspond to a {\em lower} flux than what we assume.  

An easy way to avoid this problem is to send photons along shorter
rays. Of course, this has the disadvantage of sampling smaller
portions of the IGM, possibly missing individual self-shielded
structures. It is thus optimal to choose a ray length such that the rays on
average remain optically thin, while still sampling a sufficiently long
distance to take into account the self-shielding of individual dense
structures.  We achieve this by first calculating $\lambda$ as a
function of $s$, 
and then choose the optimal ray length to be the largest
value of $s$ for which $s \leq \lambda(s) / 5$.  
A lower cutoff of $s \geq L / 32$ is also applied.  
If this condition cannot be satisfied we flag the given region of parameter
space and omit its inclusion in our analysis. For ray lengths $s < L$, 
the usage of the box is maximized by resetting the intensity
along each ray after a distance of $s$, until the ray has
traversed a distance $L$.

\begin{figure} \smjustify
\includegraphics[width=\smwidth]{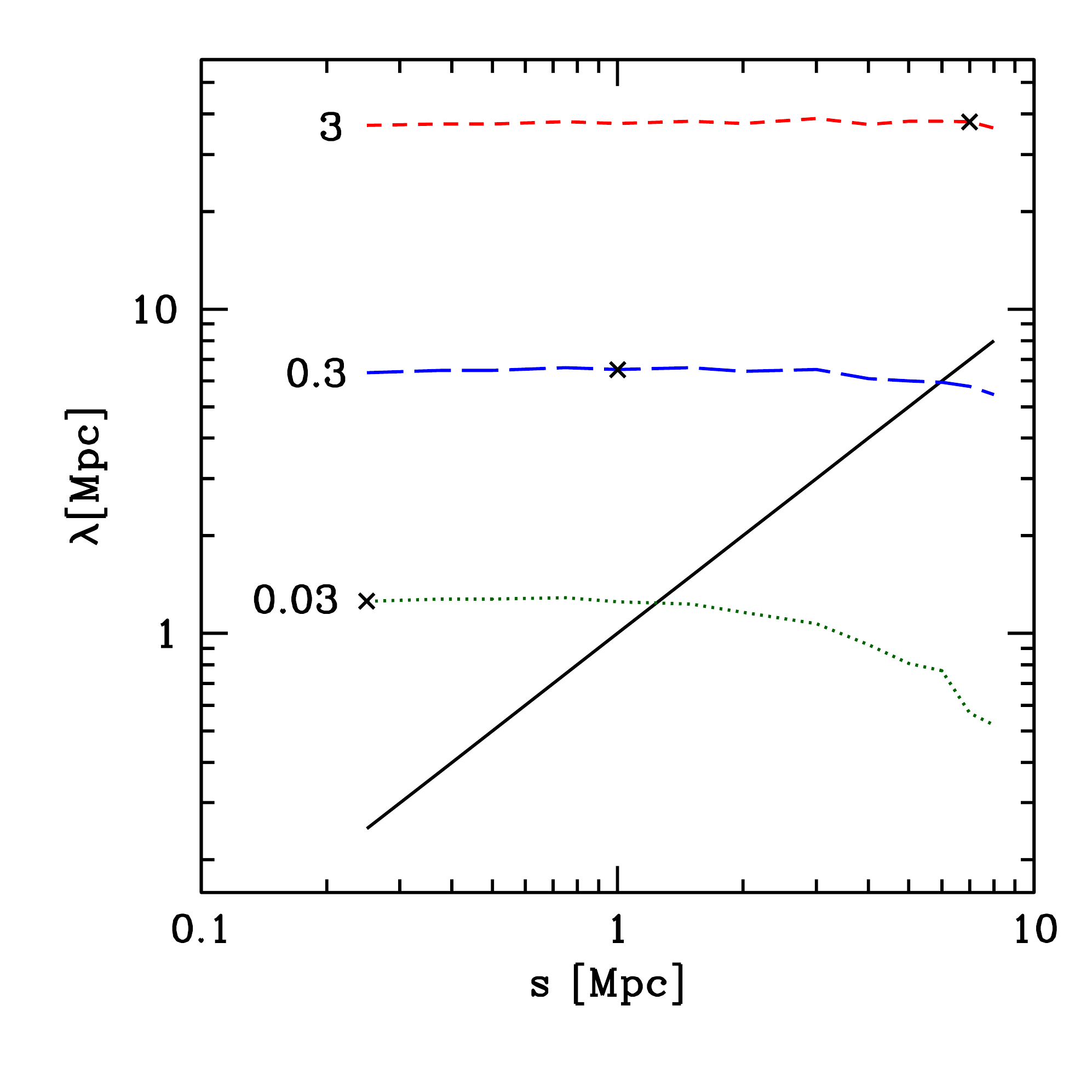} 
\vspace{-0.8cm}
\caption{Here we demonstrate our procedure for choosing the optimal
ray length that adequately samples the IGM while remaining optically
thin. The data pertains to simulation B6 ($512^3$, 8 Mpc) taken
at redshift $z = 10$. The nearly horizontal lines show $\lambda$ as
a function of ray length $s$ for different  photoionization rates
where the labels denote $\Gamma_{-12}$ in units of $10^{-12}\ {\rm
s}^{-1}$. For each of these lines, the optimal ray length is chosen as
the largest value of $s$ for which $s \leq \lambda(s)/5$ and is denoted by a
black cross. For comparison, the diagonal line traces out $\lambda =
s$ so that portions rightward of this curve belong to the optically
thick regime where the mean free path changes significantly.}  
\vspace{0.2cm}   
\label{figure:slabs}
\end{figure}

Figure \ref{figure:slabs} demonstrates our procedure of selecting
ray lengths at different fluxes for simulation B6 at $z = 10$. 
The mean free path converges in the
optically thin limit  where $\lambda \gtrsim s$ but begins to deviate
strongly during the transition to the optically thick transition when
$\lambda$ approaches $s$. It is clear from the plot that an erroneous
value for $\lambda$ would be  obtained for an improper choice of
$s$. The convergence of $\lambda$ in the optically thin limit shows that
our choice of picking $s$ to be bounded by $\lambda/5$ is robust
in changing this fraction by a factor of a few. 

\section{Simulation Results}

We first describe the dependence of clumping factor, $c_l$, of ionized
gas on redshift and photoionization rate. Next, we use the gas PDF
matched to the clumping factors we have obtained, to define a critical
overdensity, $\Delta_{\rm crit}$, above which gas remains
self-shielded  and neutral. The opacity of the IGM to ionizing
radiation, expressed in terms of the mean free path, $\lambda$, is
described next.  We first discuss its overall properties and then show
how it can be used to relate the emissivity of ionizing sources to the
photoionizing background that they produce. Finally, we compare the
clumping factors and mean free paths  obtained here to those which
would be expected for an optically thin model of the IGM. 

The main goal of this work is to assess the small-scale convergence of
numerical quantities during the initial phase of reionization. This is
presented in $\S$4 where we show simulation C3 ($1024^3$, 1 Mpc)
to be our ``converged'' simulation. However, since C3 was only evolved
to $z = 10$, we show here results from our fiducial simulation B2
($512^3$, 0.5 Mpc) in order to present results down to $z = 6$. Table
\ref{table:datapts2} summarizes the clumping factors, critical
overdensities, and mean free paths at select redshifts and
photoionization rates for simulation B2. These values  are within
$6\%$ of those from the converged simulation C3 for all $z \geq 10$. 

\subsection{Clumping Factor}

Studies of reionization typically make use of the clumping factor of
ionized gas,  defined as 
\bq
c_l \equiv \langle n_e^2 \rangle / \langle n_e \rangle^2,
\label{eq:cl}
\eq
where $n_e$ is the number density of free elections and 
angled brackets denote volume averages over space.
The clumping factor describes the enhancement of the recombination
rate relative to a uniform gas distribution, and is therefore
crucial in understanding the role of inhomogeneities in 
the ionizing photon budget during and after reionization. 

Before proceeding to discuss the clumping factors obtained here, it
is first useful to make some comments regarding
the form of equation (\ref{eq:cl}). This equation involves a volume average
over the free electron density in each ray segment of the box without applying any density
cutoffs in considering which electrons contribute to recombinations within the IGM.
We have made this choice to facilitate the use of equation (\ref{eq:clumpcrit}) 
which allows us to determine the critical overdensity required for self-shielding
within a patch of the IGM that does not contain any reionizing sources.
This can then be applied in studies that simulate self-shielding by turning
off an optically thin flux above a threshold overdensity.
Another choice would be to compute the clumping factor based only on gas
with overdensity below some cutoff that is assumed to represent the
maximum density for which recombinations occur within the IGM.
This is ideal for the scenario
where one is interested in separating recombinations occurring within the IGM
from those occurring within the interstellar medium (ISM) of ionizing sources. The 
latter can be accounted for through the use of an escape fraction describing 
the fraction of ionizing photons that escape the ISM of reionizing sources into the surrounding IGM. 
It is important to note that the numerical value obtained for the clumping factor
depends on the particular definition that is used. Regardless, the convergence tests
described in $\S$4, which are based on the clumping factor described above, remain
robust to whatever definition of clumping factor is assumed. 
For more detailed discussions of the clumping factor and comparisons between
different definitions we refer the reader to the recent works of \citet{shull/etal:2012}
and \citet{finlator/etal:2012}.

\begin{figure*} \smjustify
\includegraphics[width=\smhwidth]{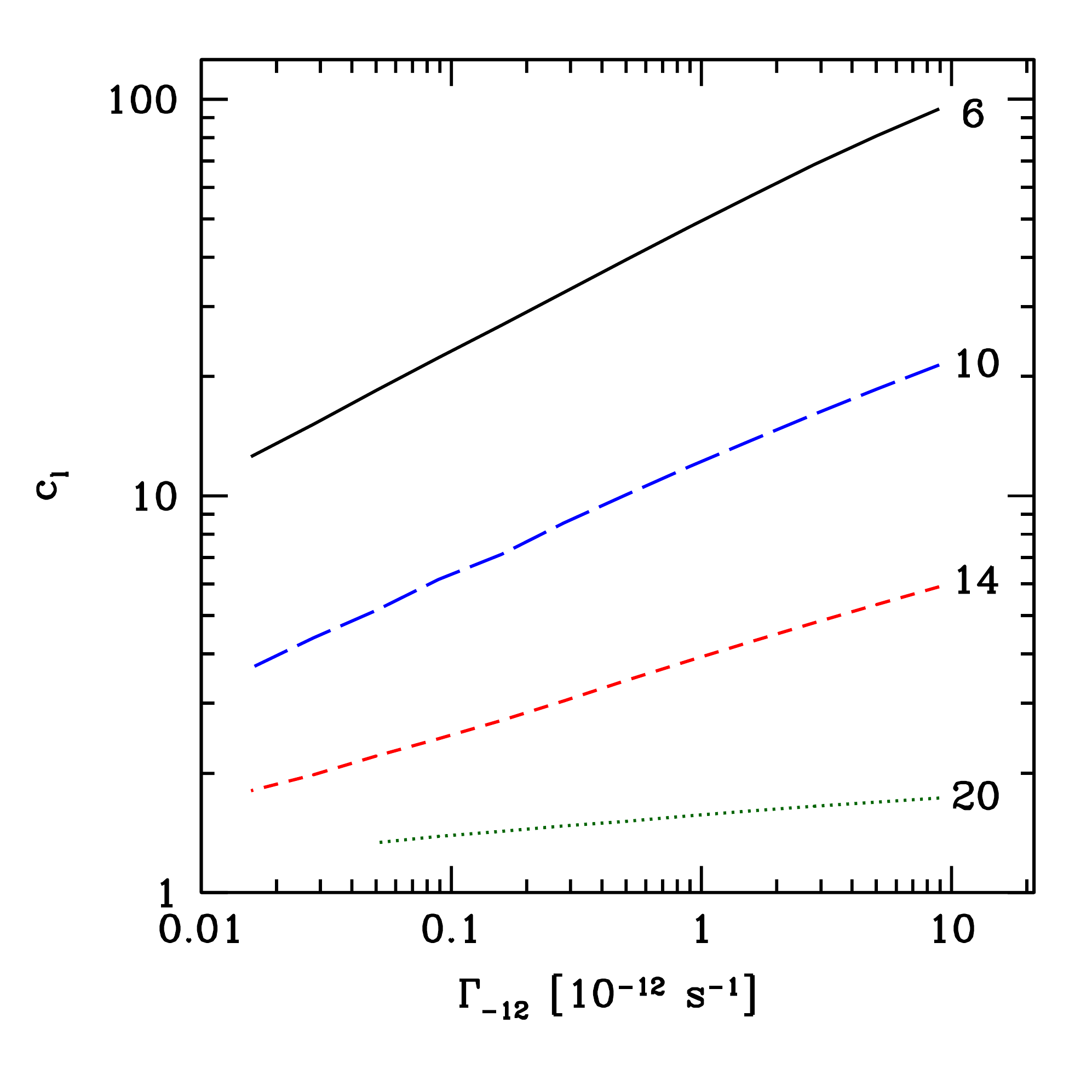}
\includegraphics[width=\smhwidth]{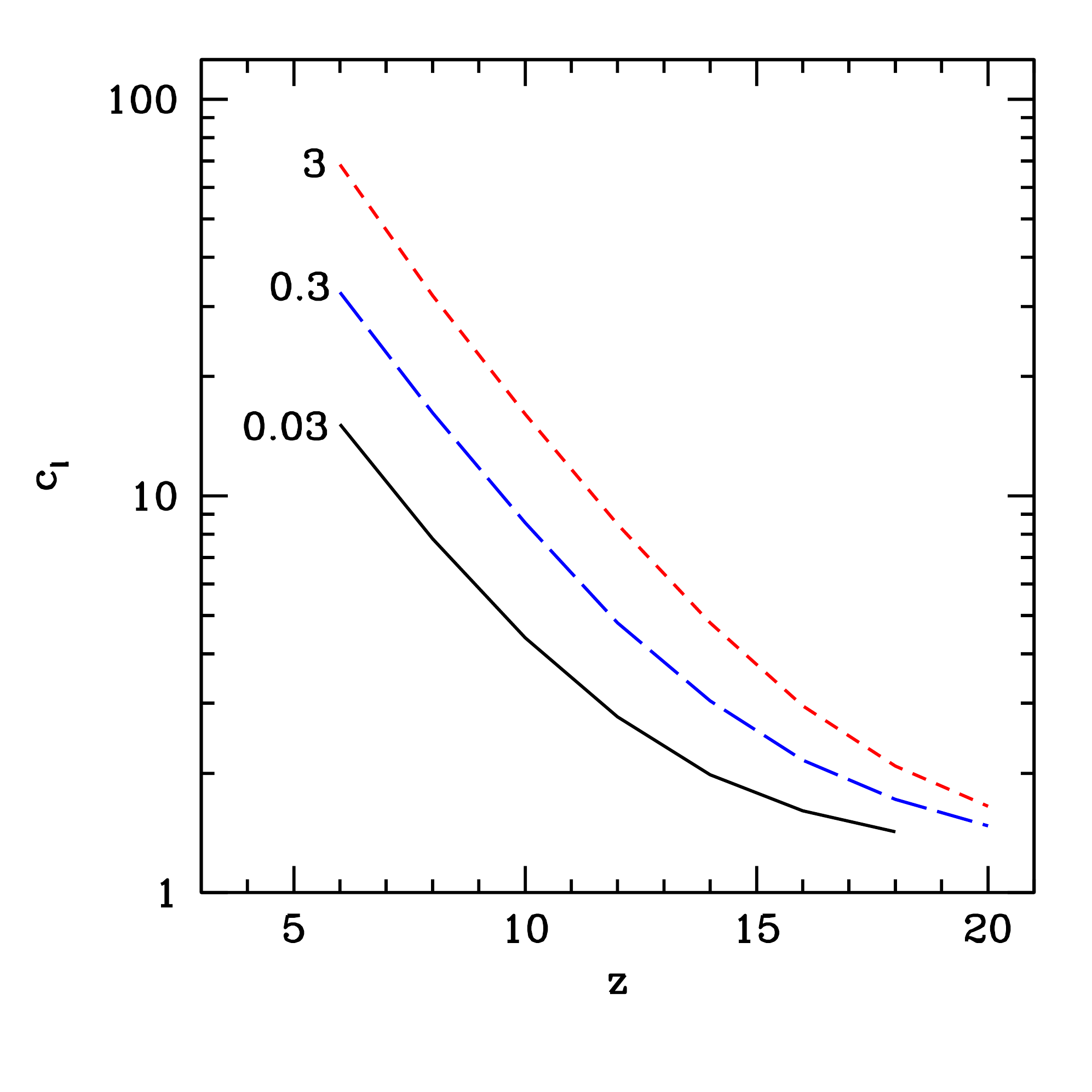}
\caption{(left) Clumping factor versus photoionization rate
  $\Gamma_{-12}$ in units of $10^{-12}~{\rm s}^{-1}$ for different
  redshifts, as labelled beside the curves. (right) Clumping factor
  versus redshift for different  photoionization rates with labels
  denoting the value of $\Gamma_{-12}$. In both cases the data
  pertains to our fiducial simulation B2 ($512^3$, 0.5 Mpc) and 
  regions of parameter space ($z > 18$ and $\Gamma_{-12}
  < 0.05$) that do not satisfy the ray length criterion described in
  $\S2.2.4$ are omitted.}
\label{figure:clumping} \vspace{0.2cm}
\end{figure*}

We now proceed to discuss the clumping factor obtained from the 
radiative transfer calculations performed on our fiducial simulation
B2. This is shown in Figure \ref{figure:clumping} where we plot 
$c_l$ for a variety of redshifts and fluxes.
Some general trends of the clumping  factor are shown
by comparing the two panels of this figure. In the first place, 
at fixed photoionization rate, $c_l$ increases with decreasing redshift
which is consistent with ongoing structure formation within the IGM.
Furthermore, at fixed redshift, $c_l$ increases with the strength of the ionizing
background as the flux of ionizing photons are able to penetrate
further into thick gas clouds, exposing their dense interiors where
the recombination rate is greatest.  Eventually, at large enough flux
of $\Gamma_{-12} \gtrsim 1000$, the clumping factor 
plateaus as the ionization state of the box saturates and all the gas 
has been ionized. At this point  $c_l$ tends to
the total clumping factor of gas in the box as $n_e$ approaches $n$.

Historically, clumping factors of $c_l \sim 30$ at $z \sim 6$ have
been found to be appropriate \citep[e.g.,][]{gnedin/ostriker:1997},
though more recently there has been a growing trend towards values an
order of magnitude smaller. It thus appears contrary to historical
development that we reproduce  
$c_l \sim 30$ at $z = 6$ with our fiducial case of $\Gamma_{-12} = 0.3$, 
and find even larger values with increased flux. However, as explained by
\citet{pawlik/etal:2009}, the passage of an ionization front through
the IGM will photoevaporate the smallest halos in the box and 
consequently suppress the evolution of the clumping factor at small
scales as the gas is dispersed back into the diffuse IGM.  Since we do
not include such hydrodynamic feedback processes in our analysis, the
values reported here cannot be used in reference to an IGM that has been
heated through photoionization. Nevertheless, our values are perfectly
applicable to the early stages of reionization, before the gas has
had time to respond to the ionizing radiation field. Moreover,
as discussed in $\S5$, previous simulations have underestimated the
clumping factor by a factor of $\sim 3$ during this period, and may 
therefore be underestimating its subsequent evolution and the impact
that unresolved small-scale structure had in regulating the early
stages of reionization. 

\begin{table}[t]
\centering
\caption{Opacity of the Unheated IGM at Select Values}
\begin{tabular}{c c c c c c}
\hline\hline
z & $\Gamma_{-12}$ $[10^{-12}\ {\rm s}^{-1}]$ & $c_l$ & $\Delta_{\rm crit}$ & $n_{\rm crit}$ [cm$^{-3}$] & $\lambda$ [Mpc]  \\ 
\hline 
18      & 0.03 & 1.4 & 6.1    & 0.008  & 0.1 \\ 
14      & 0.03 & 2.0 & 14   & 0.009 & 0.3 \\
10      & 0.03 & 4.4 & 34   & 0.008 & 0.7 \\
8       & 0.03 & 7.8 & 55   & 0.007  & 1.1 \\
6       & 0.03 & 15  & 100  & 0.006 & 2.1  \\
\hline
18      & 0.3  & 1.7 & 25   & 0.031 & 0.9  \\
14      & 0.3  & 3.0 & 52   & 0.032 & 1.6 \\
10      & 0.3  & 8.6 & 120  & 0.029 & 3.0 \\
8       & 0.3   & 16  & 200  & 0.027 & 4.5\\
6       & 0.3  & 33  & 390  & 0.025 & 8.3  \\
\hline
18      & 3    & 2.1 & 110  & 0.14 & 7.1 \\
14      & 3    & 4.8 & 210  & 0.13 & 10   \\
10      & 3    & 16  & 470  & 0.11 & 15  \\
8       & 3     & 32  & 760 & 0.10 & 21 \\
6       & 3     & 68  & 1400 & 0.09 & 36  \\
\label{table:datapts2}
\end{tabular}
\end{table}

\subsection{Critical Overdensity for Ionization}

Since the clumping factor describes the distribution of ionized gas
within the volume, it is in principle derivable from knowledge of the
gas PDF and details of the photoionizing radiation field. In a
simplified description, we assume that all gas within the box with
overdensity $\Delta < \Delta_{\rm crit}$ is ionized, while the rest is
neutral. This is obviously an idealized description of reality where
a gradual transition between ionized and neutral regions will
necessarily occur. Any departures from the simplified model reflect
variations in the local ionizing background and degree of
self-shielding and shadowing within the inhomogeneous IGM
\citep{miralda-escude/etal:2000}. 

In the simplified model the clumping factor of ionized gas is
related to the total gas PDF through the following expression: 
\bq
c_l = \frac{\int_0^{\Delta_{\rm{crit}}} \Delta^2 P(\Delta)\ d\Delta} {
\left( \int_0^{\Delta_{\rm{crit}}} \Delta P(\Delta)\ d\Delta
\right)^2}, 
\label{eq:clumpcrit} 
\eq  
where $\Delta_{\rm{crit}}$ is
interpreted as the critical overdensity above which self-shielding
prevents the gas from becoming ionized. It is often useful to assume
the form of equation (\ref{eq:clumpcrit}) taken with some nominal
choice for $\Delta_{\rm crit}$ in order to compute $c_l$ from a given
gas PDF.  For example, \citet{chiu/etal:2006} consider a model where
all gas within collapsed halos is self-shielded while all remaining
gas is subject to ionization from an UVB. In this case, $\Delta_{\rm
crit} = 6 \pi^2$  corresponding to the overdensity at the virial
radius of an isothermal sphere with a mean overdensity of 
$\Delta_{\rm vir} = 18\pi^2$.

Since we compute the clumping factor directly from our radiative
transfer calculations, we take the opposite approach, inverting
equation (\ref{eq:clumpcrit}) in order to compute $\Delta_{\rm crit}$ 
from knowledge of $P(\Delta)$ and $c_l$. In doing
so, we observe the expected trend that $\Delta_{\rm crit}$ increases when
the photoionization rate is increased, making the medium more
susceptible to ionization. In fact, we find the rough
proportionality $\Delta_{\rm crit} \propto \Gamma_{-12}^{2/3}$ which,
from equation (4) of \citet{mcquinn/etal:2011}, is expected for
a PDF satisfying $P(\Delta) \propto \Delta^{-2.5}$. We showed in
Figure \ref{figure:pdfs} that our PDFs satisfy this power-law at $z \lesssim 10$
for $\Delta > \Delta_{\rm vir}$. This is consistent with the model where
gas at these densities is collapsed in isothermal spheres. 
Around our fiducial value of 
$\Gamma_{-12} = 0.3$, we further find that $\Delta_{\rm crit}$ is roughly 
proportional to $(1+z)^{-3}$, indicating that the critical proper
hydrogen number density, $n_{\rm crit}$, is rather insensitive to redshift.
A good value to take is 
$n_{\rm crit} \sim 0.1$~cm$^{-3}~\Gamma_{-12}^{2/3}$.

\begin{figure} \smjustify
\includegraphics[width=\smwidth]{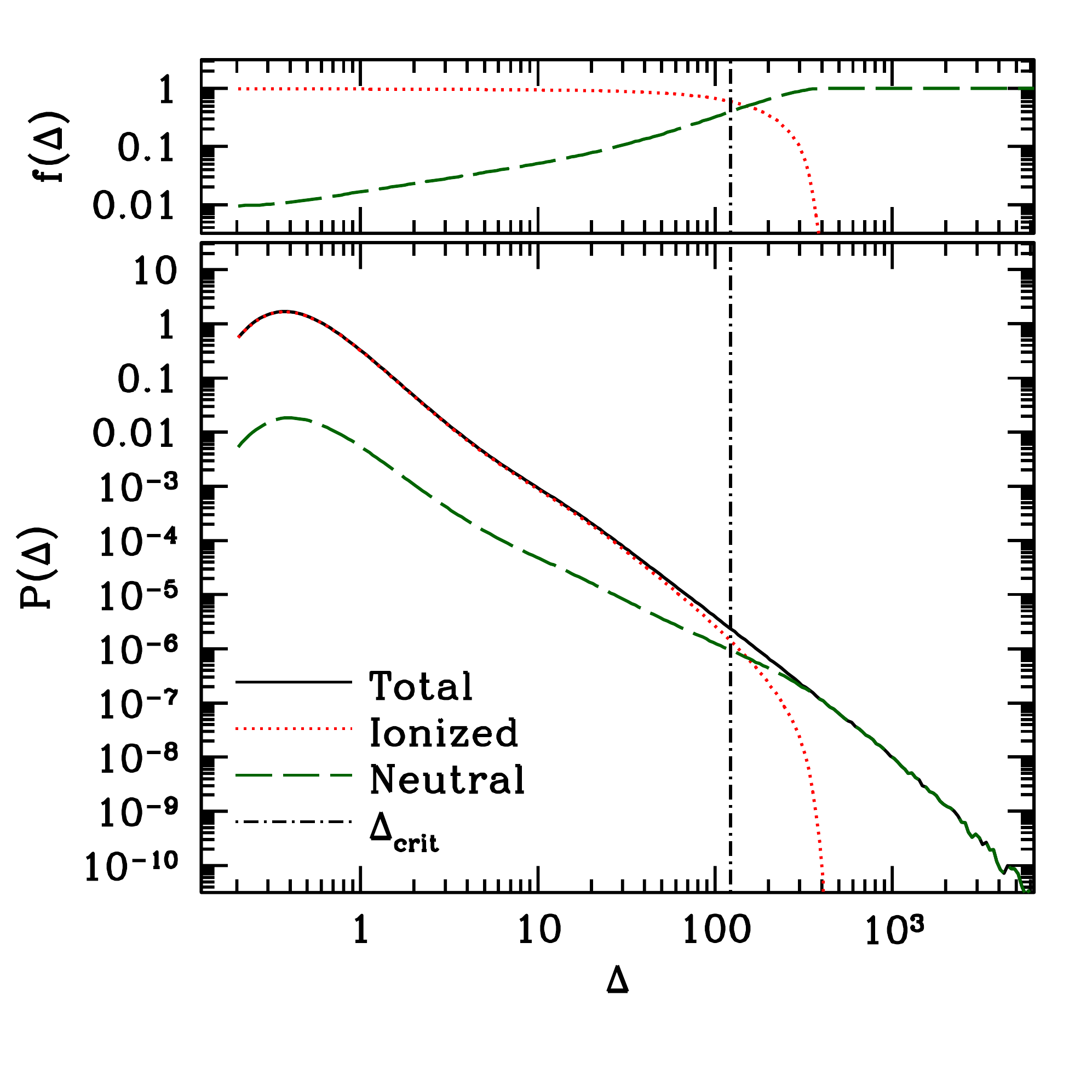} \vspace{-0.7cm}
\caption{The bottom panel compares the total gas PDF of our fiducial simulation 
B2 ($512^3$, 0.5 Mpc) to the PDFs of neutral and ionized gas
within the box after radiative transfer is applied.  Ray segments are
labelled as ionized if they have an ionized fraction greater than 0.5
and are labelled neutral otherwise. The data corresponds to a
snapshot at $z = 10$ with an ionizing background  of $\Gamma_{-12} =
0.3$. The vertical dot-dashed line denotes $\Delta_{\rm crit}$
determined by  comparing $c_l$ obtained from the radiative transfer
calculation to the total gas PDF through  equation
(\ref{eq:clumpcrit}). The top panel shows the  corresponding volume
fraction of neutral and ionized gas as a function of $\Delta$.} 
\label{figure:deltacrit} \vspace{0.2cm}
\end{figure}

The validity of the idealized model where all gas with overdensity
$\Delta < \Delta_{\rm crit}$ is ionized is tested in Figure
\ref{figure:deltacrit}. Here we plot the total gas PDF along with
the ionized and neutral PDFs obtained from our radiative transfer
calculation using our fiducial parameters $\Gamma_{-12} = 0.3$ and
$z = 10$ for simulation B2. 
The vertical dot-dashed line shows the corresponding value of
$\Delta_{\rm crit}$ -- its role in delineating the neutral and
ionized portions  of the gas is clearly visible. As anticipated,
the transition between ionized and neutral regions is not sharp, but
rather gradual as a consequence of the spatially varying ionizing
background and self-shielding due to dense gas pockets. Nevertheless,
our findings indicate that the approximation that $\Delta_{\rm crit}$
represents the critical overdensity above which self-shielding maintains
the neutral state of the IGM is generally a good one.

\subsection{Mean Free Path}

We quantify the opacity of the IGM to the exposed UVB through the use
of the mean free path of ionizing radiation. Conceptually, one can 
consider the mean free path to be affected by two components: a diffuse
gaseous phase that pervades the IGM and thick gas clouds embedded 
within collapsed dark matter halos. The latter 
make up a significant fraction of absorption systems that have
neutral hydrogen column densities $N_{HI} \gtrsim 1/\bar{\sigma}
\approx 10^{17}\ {\rm cm}^2$ allowing them to self-shield against
ionizing radiation. It is within these optically-thick structures
where the global recombination rate of the IGM is dominated and the
majority of ionizing photons are absorbed
\citep{miralda-escude/etal:2000}. As a result, they can significantly
impede the progress of reionization.

\begin{figure*} \smjustify
\includegraphics[width=\smhwidth]{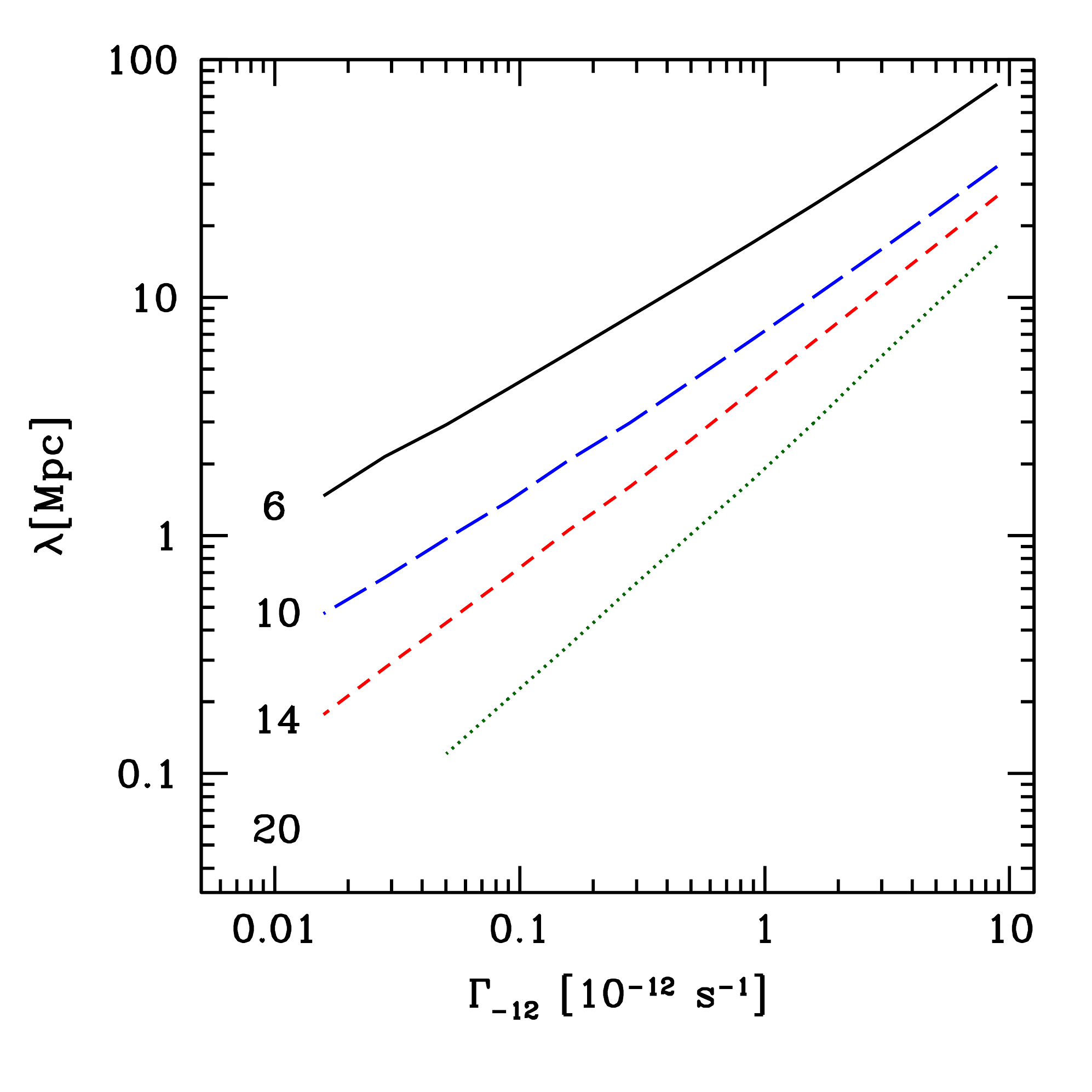}
\includegraphics[width=\smhwidth]{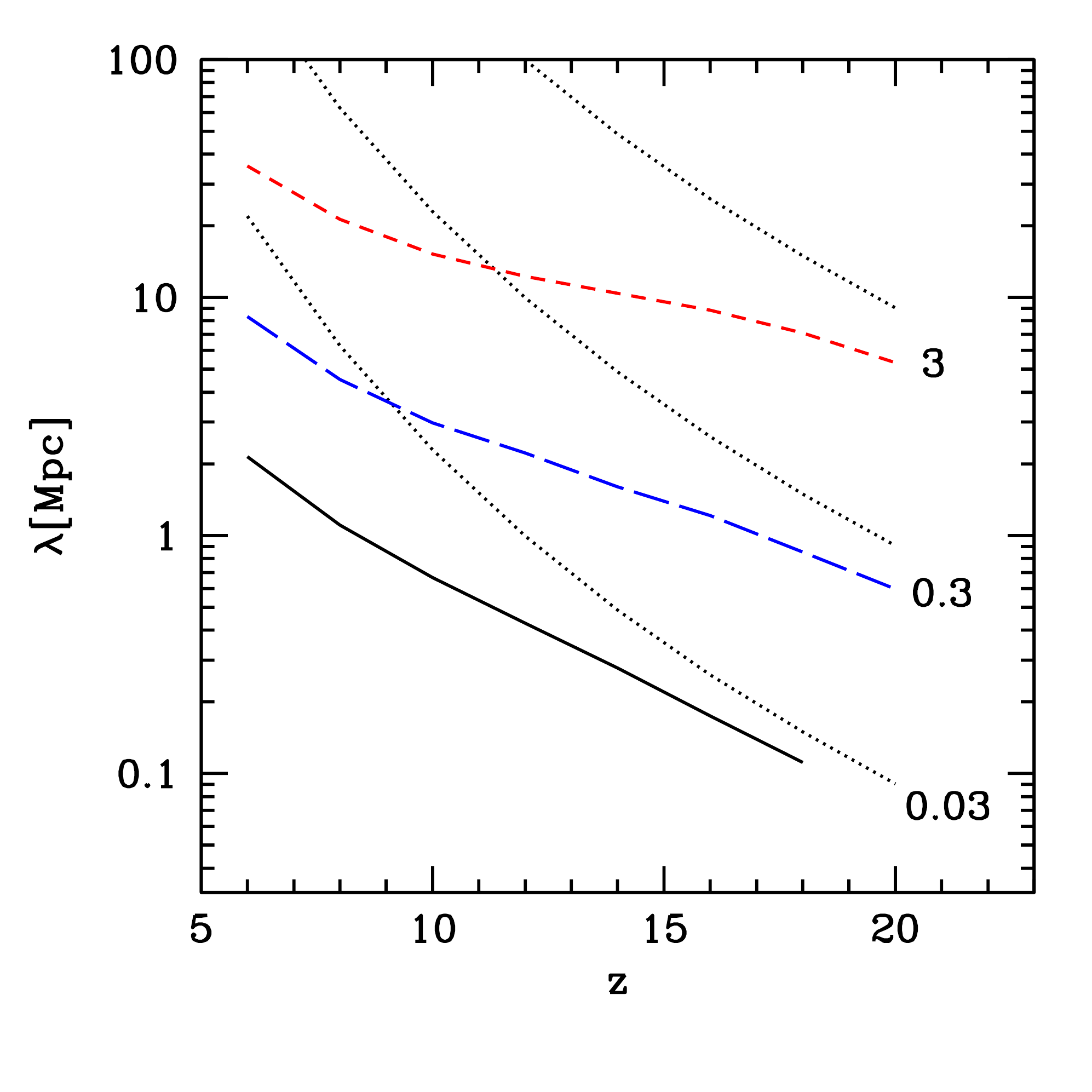}
\caption{(left) The mean fee path versus photoionization rate $\Gamma_{-12}$ in
units of $10^{-12}~{\rm s}^{-1}$ for different redshifts, as labelled.
(right) Mean free path versus redshift for different photoionization
rates with labels indicating the value of $\Gamma_{-12}$.
In both cases the data pertains to our fiducial
simulation B2 ($512^3$, 0.5 Mpc) and
 regions of parameter space ($z > 18$ and $\Gamma_{-12}
  < 0.05$) that do not satisfy the ray length criterion described in
  $\S2.2.4$ are omitted.
The dotted black curves in the right panel 
show the mean free path expected for
an optically thin, completely ionized, and homogeneous IGM as expressed in equation (\ref{lambdamean}).
From bottom to top in the plot, the dotted lines take $\Gamma_{-12} = 0.03$, $0.3$, and
$3$, and are each calculated using $x = c_l = 1$. We expect the dotted lines 
to converge with our results at high redshift when the medium approaches homogeneity. 
At low redshift we observe a large suppression in the calculated mean free path that
results from increased structure formation within the inhomogeneous IGM. 
}
\label{figure:lambda} \vspace{0.2cm}
\end{figure*} 

The mean free path obtained from our radiative transfer calculations
is computed through  the use of equation (\ref{eq:slabmfp}) which
naturally encompasses both the clumpy IGM and halo components. In
Figure \ref{figure:lambda} we plot the mean free path as a function
of photoionization rate for fixed redshift and also as a function of
redshift for fixed photoionization rate.
Some general trends are immediately clear in this plot. In the
first place, at fixed redshift we see that the mean free path
increases with the strength of the ionizing background. A  stronger
flux of ionizing photons will naturally penetrate further through a
diffuse IGM and overcome  thicker self-shielding structures,
consistent with the previous observation that  $\Delta_{\rm crit}
\propto \Gamma_{-12}^{2/3}$. In addition, when the ionizing background is
held constant, the mean free path is found to increase with decreasing
redshift. 

It is important to note that there are two competing factors affecting
the redshift evolution of $\lambda$.  On the one hand, the expansion
of the Universe continually dilutes the density of hydrogen, hence
favouring  a strong increase in $\lambda$ with decreasing $z$. On the other hand,
increased structure formation at low redshift enhances the
distribution of Lyman-limit systems that strongly inhibit the distance
an ionizing photon can  propagate through the IGM before being
absorbed.  In the right panel of Figure \ref{figure:lambda} we compare
$\lambda$ obtained here to equation (\ref{lambdamean})  for the same
set of photoionization rates. Taking $x = c_l = 1$ in this equation
yields the mean free path we would obtain in an optically thin,
homogeneous, and completely ionized medium. In such a model the mean
free path evolves rapidly with redshift as $\lambda \propto
(1+z)^{-5}$. Instead, we observe a strong suppression in mean  free
path at low redshift compared to equation (\ref{lambdamean}). This
highlights the important contribution from inhomogeneities in the
IGM. 

\subsection{Relationship Between Emissivity and Photoionization Rate}

\begin{figure} \smjustify
\includegraphics[width=\smwidth]{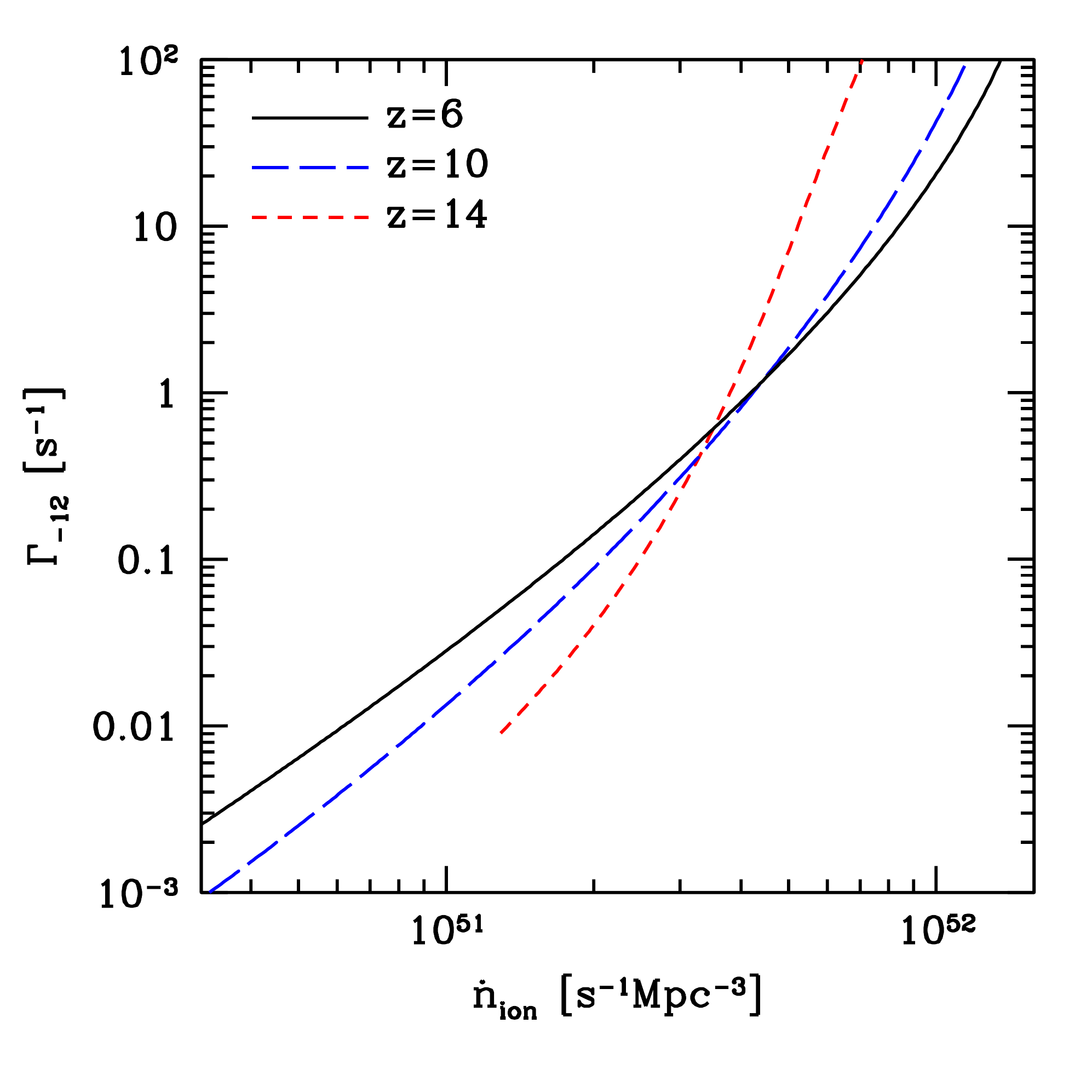} \vspace{-0.7cm}
\caption{The photoionization rate, expressed in terms of
$\Gamma_{-12}$, as a function  of the comoving ionizing emissivity at
different redshifts. These curves are evaluated by solving equation
(\ref{eq:ndot}) using $\lambda(\Gamma, z)$ obtained from our radiative
transfer calculations performed on our fiducial simulation B2
($512^3$, 0.5 Mpc).  At low flux we truncate the $z = 14$ curve when
it enters the optically thick domain where  it no longer satisfies the
ray length criterion described in $\S2.2.4$.}
\label{figure:ndot} \vspace{0.2cm}
\end{figure}

In the context of reionization it is desirable to know the ionizing
background produced by some population of sources with known
emissivity.  This relationship can be found by solving 
\bq 
\Gamma = \dot{n}_{\rm ion} (1 + z)^2 \lambda(\Gamma, z)
\bar{\sigma},
\label{eq:ndot} 
\eq 
where $\dot{n}_{\rm ion}$ is the comoving ionizing
emissivity and $\lambda( \Gamma, z)$ is the comoving mean free path
that depends on both the ionizing background and redshift. In Figure
\ref{figure:ndot} we show the dependence of $\Gamma$ on
$\dot{n}_{\rm ion}$ by solving equation (\ref{eq:ndot}) with
the mean free paths taken from our radiative transfer calculations. 
We find that  $\Gamma$ exhibits a rather steep
dependence on emissivity and appears to diverge at large values of
$\dot{n}_{\rm ion}$. This behaviour is attributed to the fact that not
only are there more ionizing photons as the emissivity
rises, but also their ability to penetrate further through the IGM increases. 

\begin{table}[b]
\centering
\caption{Power law index of $\Gamma\propto\dot{n}_{\rm ion}^\gamma$}
\begin{tabular}{c | c c c c c c c c}
\hline\hline
 & \multicolumn{7}{c} z \\
  & 6 & 8 & 10 & 12 & 14 & 16 & 18 & 20 \\
\hline
$\gamma$ & 2.6 & 2.9 & 3.1 & 3.7 & 4.5 & 6.7 & 10 & 14
\label{table:beta}
\end{tabular}
\end{table}

We can relate this behaviour back to the dependency of $\lambda$ on
$\Gamma$. For instance, suppose we have the simple relation $\lambda
\propto \Gamma^\beta$ at some redshift.  Then from equation
(\ref{eq:ndot}) we will have that $\Gamma \propto \dot{n}_{\rm
ion}^\gamma$  where $\gamma = (1-\beta)^{-1}$. In Table
\ref{table:beta} we list the values of $\gamma$ obtained by fitting a
power-law to our fiducial mean free paths within the range $0.1\leq
\Gamma_{-12} \leq 1$ at different redshifts.  This flux range is
considered to emphasize the relationship between $\Gamma$ and
$\dot{n}_{\rm ion}$ around our fiducial value of $\Gamma_{-12} = 0.3$.
We find that the relationship between $\Gamma$ and $\dot{n}_{\rm ion}$
strengthens as the redshift increases -- $\gamma$ varies from 2.6 to
14 between redshifts 6 and 20 respectively.  This occurs because the
slope $\beta$  rises as the IGM becomes more uniform, approaching a
limiting value of unity for a completely homogeneous Universe with
$c_l = 1$ in equation (\ref{lambdamean}) -- a manifestation of
``Olber's Paradox''.

From this trend we can deduce that decreasing the simulation
resolution should steepen the  curves in Figure \ref{figure:ndot} as
the density distribution becomes more homogenous.  Indeed, this
relation is observed between our suite of simulations where we find
$\gamma = 3.2$  at $z = 6$ for our worst-resolved simulation A6
($256^3$, 8 Mpc), compared to $\gamma = 2.6$  for simulation
B2. \citet{mcquinn/etal:2011} report the value of $\gamma \sim 4$ at
$z = 6$.  The discrepancy with our result likely arises from a
combination of our increased resolution and our omission of
photoheating which would suppress accretion  of gas onto low-mass
halos and promote homogeneity. 

The photoionization rate after reionization can be derived from
measurements of the Ly$\alpha$ forest. This was done by
\citet{kuhlen:2012} who list the values of  $\Gamma$ and
$\dot{n}_{\rm ion}$ for redshifts between 2 and 6. The quoted values
at $z = 6$ are $\Gamma_{-12} < 0.19$ and $\dot{n}_{\rm ion} < 2.6
\times 10^{50}~{\rm s^{-1}~ Mpc^{-3}}$. Looking at Figure
\ref{figure:ndot} we see that our $z = 6$ curve predicts an
emissivity an order of magnitude too large when $\Gamma_{-12} = 0.19$. 
As we discuss in $\S5$, our results at
$z \lesssim 10$ are hindered by the omission of radiative feedback
that would otherwise smooth inhomogeneities on small
scales. 
Photoheating suppresses structure growth within the IGM thereby
increasing the mean free path at fixed conditions as ionizing photons
travel further before encountering an optically thick absorption system.
From equation (\ref{eq:ndot}) a larger mean free path produces a higher
ionization rate at fixed emissivity, implying that the $z = 6$ curve in Figure  
\ref{figure:ndot} would shift upwards in the presence of heating. 
If this shift were to bring us into agreement with observations we would
require that our calculated mean free path at $\Gamma_{-12} = 0.19$ 
increase by an order of magnitude to allow for an order of magnitude 
reduction in the emissivity required to produce such a background. 
In Figure \ref{figure:lambda} we find 
$\lambda \approx 7$ Mpc at $z = 6$ for $\Gamma_{-12} = 0.19$. 
This means that a mean free path of 70 Mpc would be required in a {\em heated}
IGM to bring us into agreement with observations of the Ly$\alpha$ forest.
This is in reasonable agreement with the value of $\lambda = 49 \pm 14$ Mpc
reported by \citet{kuhlen:2012} as the mean free path at the Lyman edge.
Note that at redshifts $z \gtrsim 10$, before photoheating is
important, the curves in Figure \ref{figure:ndot} should be correct.
Of course, reionization becomes patchy at high redshift, making a 
description in terms of an IGM with a single UVB flux and mean free 
path less accurate.

The strong scaling relations observed here suggest that small changes
in $\dot{n}_{\rm ion}$ can  boost $\Gamma$ by substantial
amounts. \citet{mcquinn/etal:2011} use this to argue that the rapid
evolution in $\Gamma$ observed by \citet{Fan06} at $z \approx 6$ can
be explained by a small change in the emissivity of the ionizing
background rather than attributing this effect to the overlap phase
of reionization. 

\subsection{Relationship Between Mean Free Path and Clumping Factor}

Suppose there is an infinitesimally thin slab of width $ds$ whose area
element $dA$ is exposed to some flux $F$. In ionization equilibrium,
the number of ionizations occurring per unit time balance the number
of recombinations:   
\bq
dF dA = \alpha_B n_e n_{HII} dA ds,
\label{eq:ot1}
\eq
where $dF$ is the attenuation of flux passing through the
slab. Dividing both sides of this expression by $dA ds$, substituting
equation (\ref{ftransfer}), and taking $n_e = n_{HII}$ implies
\bq
F(1+z)/\lambda = \alpha_{\rm B} n_e^2.
\label{eq:ot2}
\eq
Finally, taking the clumping factor defined in equation (\ref{eq:cl})
and the ionized fraction $x \equiv \langle n_e \rangle / \langle n_H
\rangle$ we obtain: 
\bq
F(1+z)/\lambda = c_l \alpha_{\rm B} x^2 \langle n_H \rangle ^2.
\label{eq:ot3}
\eq
For an ionized gas temperature of $T_{\rm gas} = 10^4$~K,
\bq
\lambda\simeq 
\frac{23~{\rm Mpc}}{x^2 c_l}
\left(
\frac{\Gamma_{-12}}{0.3~10^{-12}~{\rm s}^{-1}}
\right)
\left(
\frac{1+z}{11}
\right)^{-5},
\label{lambdamean}
\eq
where we have made use of equation (\ref{eq:gamma12}) for the
conversion from flux to photoionization rate. 

\begin{figure} \smjustify
\includegraphics[width=\smwidth]{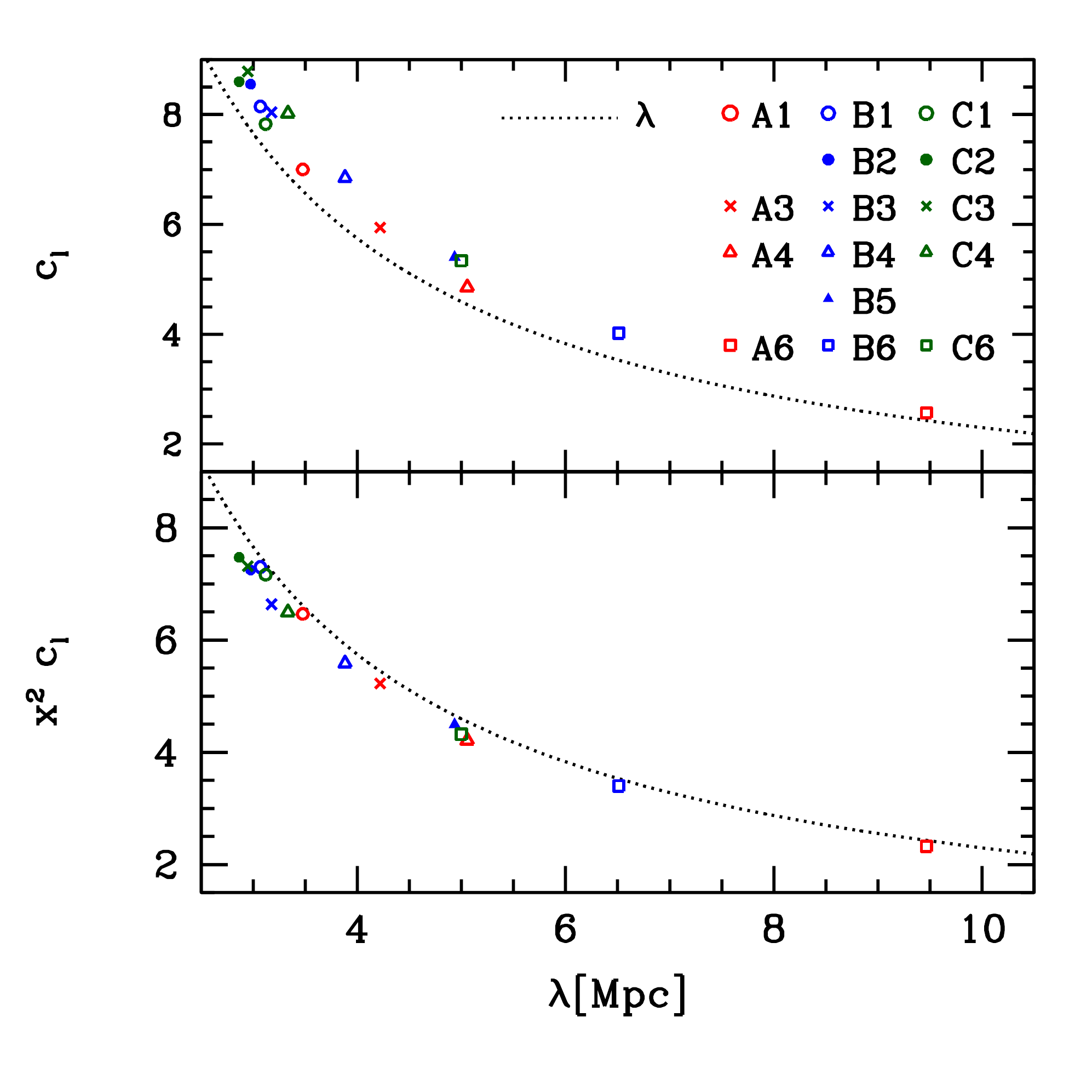} \vspace{-0.7cm}
\caption{$c_l$ versus $\lambda$ (top panel) and $x^2 c_l$ versus
$\lambda$ (bottom panel) for all simulations taken at $z = 10$ with
a photoionization rate $\Gamma_{-12}=0.3$. In each panel points denote
simulation values according to the legend in the top right corner of the plot.
Traced by the dotted line is equation (\ref{lambdamean}) which describes the
relationship between mean free path and clumping factor for an optically
thin IGM.}
\label{figure:lamclump} \vspace{0.2cm}
\end{figure}

In Figure \ref{figure:lamclump} we plot $c_l$ versus $\lambda$ and
$x^2 c_l$ versus $\lambda$ from our radiative transfer calculations at
$z = 10$ with $\Gamma_{-12} = 0.3$ for each of the simulations listed
in Table \ref{table:simparams}. In each panel the dotted line traces
equation (\ref{lambdamean}) that we would expect for an optically thin
IGM exposed to the given flux. The overall agreement between
$\lambda$ and the simulation points in the bottom panel
indicates consistency in the definition of the mean free path and
detailed balance between absorptions and ionizations. The minor
deviations between the data points and the dotted curve arise 
because our rays have a finite length, $s$, so the mean free path 
evaluated in equation (\ref{lambdamean}) will not correspond exactly
to equation (\ref{eq:slabmfp}), which is strictly correct only in the
limit where $s\longrightarrow 0$.

The simulations span a large range of values in Figure
\ref{figure:lamclump} with $c_l$ ranging from 2.6 to 8.8 and $\lambda$
from 2.9 to 9.5 Mpc. This spread arises from the broad variation in
spatial and mass resolution exhibited by the suite of simulations. In
spite of this, there is a clear  grouping of points at $c_l \sim 8$
and $\lambda \sim 3$ Mpc.  This reflects the trend towards numerically
converging to the ``correct'' clumping factor and mean  free path and
is our next topic of focus. 

\section{Numerical Convergence}

In this section, we attempt to answer the following question: {\em
What mass resolution and box sizes are necessary in cosmological
hydrodynamics simulations, in order to obtain accurate results for the
inhomogeneity of an unheated IGM?} Clearly, simulations must have
sufficient mass resolution to resolve the internal structure of the
lowest mass halos that can contain gas. In addition, however, such
simulations must cover a large enough volume to contain a
representative sample of the low-mass halos that dominate the opacity
of the IGM. 

\begin{figure} \smjustify
\includegraphics[width=\smwidth]{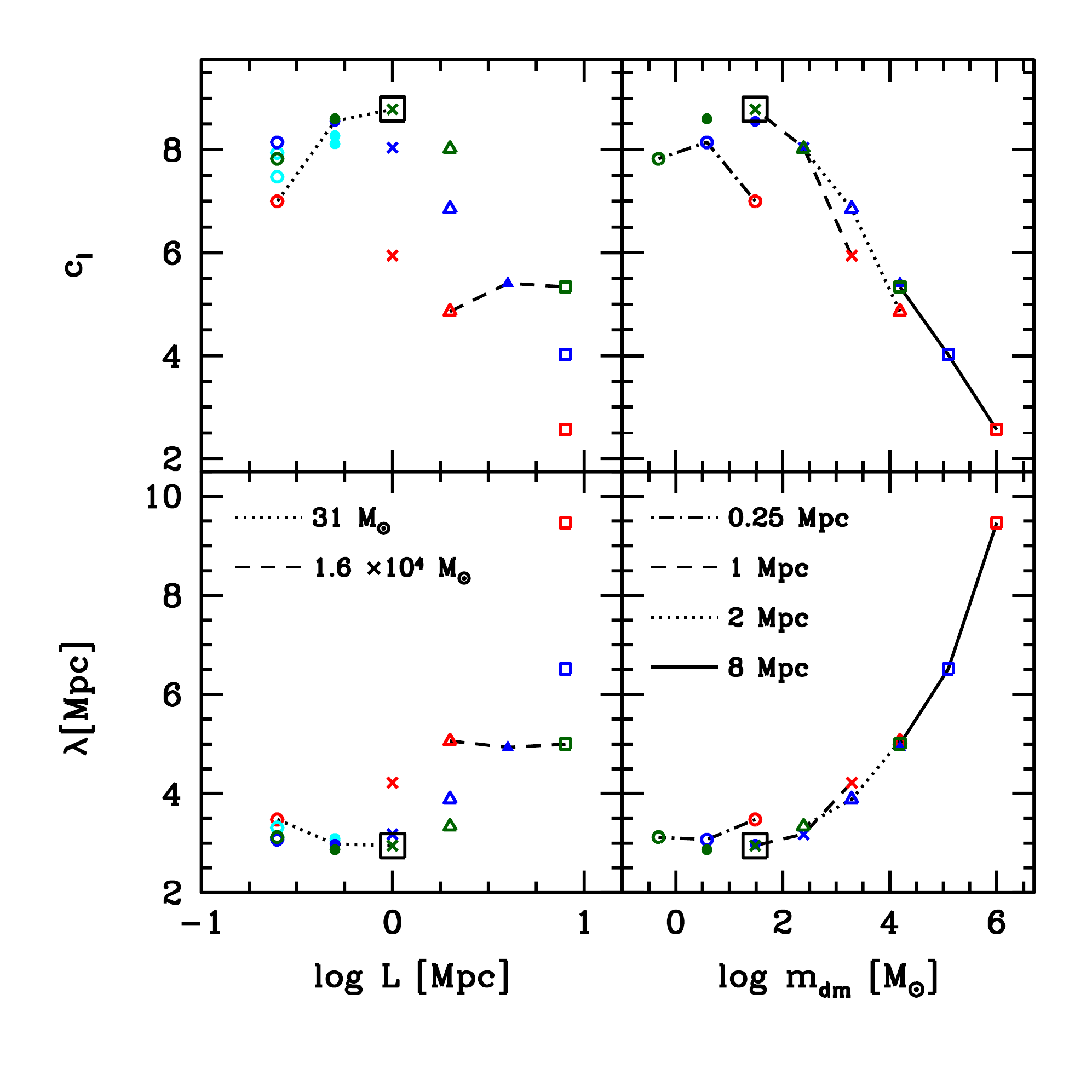} \vspace{-0.7cm}
\caption{Clumping factor and mean free path versus simulation box size (left panels)
and dark matter particle mass (right panels) for $z = 10$ and $\Gamma_{-12} =
0.3$. In each panel point types distinguish between different simulations 
and follow the legend in Figure \ref{figure:lamclump}. Lines in the left 
panels connect simulations with fixed particle mass, while lines in the right
panels connect simulations with fixed volume. The converged simulation
C3 ($1024^3$, 1 Mpc) is denoted by a green cross and is highlighted by a black square for 
easy identification. In the two left panels, open cyan circles and closed 
cyan circles denote the values of $c_l$ and $\lambda$ obtained from
different random realizations of simulations B1 and B2 respectively.}
\label{figure:conv} \vspace{0.2cm}
\end{figure}

In Figure \ref{figure:conv} we plot the clumping factors and mean free
paths  obtained from the radiative transfer calculations performed on
each of the simulations  listed in Table \ref{table:simparams}. To
obtain a picture of convergence we display how $c_l$ and  $\lambda$
vary as functions of simulation box size, $L$, (left panels) and dark
matter particle mass, $m_{\rm dm}$, (right panels) for our fiducial
redshift $z = 10$ and photoionization rate $\Gamma_{-12} = 0.3$.
Though the discussion below pertains explicitly to these fiducial values
we have checked that the picture remains consistent for 
$10 \leq z \leq 20$ and $0.01 \leq \Gamma_{-12} \leq 10$.

Starting in the top right panel of Figure \ref{figure:conv} we show
how $c_l$ changes as the mass resolution of the simulations is
varied. As expected, the clumping factor increases as the resolution
is refined and begins to plateau to a common value of $c_l \sim 8.8$
at $m_{\rm dm} \sim 30~M_\odot$, when a sufficiently low particle mass
required to resolve the smallest gaseous structures in the box is reached.
Simulations with larger particle mass 
are unable to resolve the smallest inhomogeneities contributing to the 
clumpiness of the IGM and consequently imply clumping factors up to a
few times smaller than the converged result. In the opposite limit we find that
simulations with $m_{\rm dm} < 5~M_\odot$ are also yielding
smaller clumping factors of $c_l \sim 8$.  This appears
counterintuitive at first glance since these simulations should have
no problem resolving the Jeans scale of the IGM. However, their
inability to converge on $c_l$ is attributed to their small box size,
as explained below. The three smallest simulations with $L = 0.25$ Mpc
are connected by a dot-dashed line and all display conspicuously 
unconverged values.  

The dependence of clumping factor on box size is shown in the top left
panel of Figure \ref{figure:conv}.  In this case we expect to find a
minimum box size above which simulations converge on
$c_l$. Smaller boxes fail to capture power from large-scale modes 
and should therefore have reduced clumping factors. Larger boxes that
are able to capture a representative collection of absorption systems within
their volume should converge provided that they have sufficient mass resolution.
These general trends can be identified by the behaviour shown in the top left 
panel of Figure \ref{figure:conv}. There is a clear convergence of points near
$L \sim 1$ Mpc and $c_l \sim 8.8$ with $c_l$ falling off on either side. 
Simulations with $L \lesssim 1$ Mpc have insufficient volumes to converge
with this result, while those with $L \gtrsim 1$ Mpc have insufficient mass resolutions. 
Convergence occurs in the middle ground where both a sufficient volume and 
mass resolution are attained. 

We find similar behaviour by comparing how $\lambda$ changes between simulations.
From the bottom panels of Figure \ref{figure:conv} we see that 
the resultant behaviour is essentially an inversion of that described for
$c_l$. Firstly, simulations with coarse mass resolutions overestimate $\lambda$.
These runs are unable to resolve small-scale inhomogeneities and the degree of
self-shielding that would otherwise inhibit the  propagation of
ionizing photons through neutral patches of the IGM. Secondly, the smallest
boxes also produce values of $\lambda$ that are too large. As mentioned above,
these simulations underproduce the collection of halos that shield against the 
propagation of an ionization front through the IGM.

The simulations shown here have relatively small volumes in a
cosmological context.  One issue that must be considered with these
small boxes is that of sample variance.  In the left panels of Figure
\ref{figure:conv} we plot $c_l$ and $\lambda$ obtained from
simulations B1 ($512^3$, 0.25 Mpc) and B2 ($512^3$, 0.5 Mpc) that were
run using two  different random realizations of the same initial
density field. The clumping factors over all  three random
realizations vary by $9\%$ and $5\%$ for B1 and B2 respectively,
indicating  that sample variance is somewhat important within these
volumes. This may explain the  unexpected result that $c_l$ for
simulation C1 ($1024^3$, 0.25 Mpc) is smaller than that of simulation
B1. Normally, at fixed volume, increasing the mass resolution should
enhance the clumping factor (e.g., as seen by comparing $c_l$ between
simulations A3, B3, and C3, which are connected by the dashed curve in the top right
panel of Figure \ref{figure:conv}). However, we may not expect to observe
this trend with only one realization of a small box with large sample
variance, and must also keep this in mind when interpreting the
results of our convergence test. 

The above results indicate that convergence in $c_l$ and $\lambda$ is
attained by simulation C3 ($1024^3$, 1 Mpc) with $m_{\rm dm} =
31~M_\odot$.  This simulation has a fine enough mass resolution to
resolve small-scale inhomogeneities, and has a large enough volume
that sample variance should be unimportant and large-scale modes
should be captured.  In order to make this claim more rigorous we would
have to compare against a box with larger volume and finer particle
mass than currently feasible. Nevertheless, the data presented in Figure
\ref{figure:conv} provides compelling evidence that numerical
convergence is being approached, and we suspect that deviations in our
values for $c_l$ and $\lambda$ from their ``true'' values are small enough
to claim convergence in simulation C3.
Based on this we find that the necessary requirements for
describing the inhomogeneity of an unheated IGM using cosmological
hydrodynamics simulations is to use box sizes $L \gtrsim 1$ Mpc with
dark matter particle masses $m_{\rm dm} \lesssim 50~M_\odot$. Smaller
boxes are troubled by sample variance while coarser mass resolutions
are unable to resolve the mass scale where gaseous halos are
dominating the opacity of the IGM.

\section{Discussion}

We have performed high-resolution, cosmological simulations of
structure formation at redshifts $z>6$, including adiabatic
hydrodynamics. By post-processing the resulting density fields with a
radiative transfer algorithm for hydrogen ionizing radiation, we have
determined the opacity of the unheated IGM, in terms of the mean free
path to ionizing radiation, $\lambda$, as a function of redshift and
ionizing background intensity. These results are relevant (1) as
converged solutions for the opacity of the IGM early in the
reionization process, before photoheating has evaporated small-scale
structure and (2) in determining what mass and length resolutions are
necessary to correctly model the propagation of ionization fronts into
the neutral IGM. We derive values of $n_{\rm crit}$, the proper
hydrogen number density above which gas remains neutral, that are for
the most part a function of only $\Gamma_{-12}$. Simulations that
mimic the effect of self-shielding by turning off the optically-thin
flux at high densities should use $n_{\rm crit} \sim
0.1$~cm$^{-3}~\Gamma_{-12}^{2/3}$, independent of redshift.

Our post-processing approach neglects the hydrodynamic feedback of
photoheating on the density evolution. These results therefore indicate
what the initial degree of inhomogeneity should be as ionization
fronts propagate into the IGM. In addition, they place an upper limit
to this inhomogeneity in patches of the IGM that have already been
ionized. We find that the initial clumping factor of the IGM just as
it is being ionized is a strong function of redshift and ionizing background
intensity, with typical values at $z=10$ ranging from about $c_l=4.4$ to $16$
and $\lambda=0.7$ to $15$~Mpc, for $\Gamma_{-12}=0.03$ to
$\Gamma_{-12}=3$, respectively. 

Modelling the transition from a neutral to ionized IGM requires
self-consistent simulations of the coupled radiative transfer and
hydrodynamical photoevaporation process. \citet{shapiro/etal:2004}
used idealized two-dimensional radiative transfer hydrodynamics
calculations of the photoevaporation of initially spherical, isolated
minihalos, surrounded by infalling gas. Those calculations showed that
smaller minihalos are photevaporated faster, and that larger fluxes
lead to faster photoevaporation times as well. 
\citet{2005MNRAS.361..405I} extended these models to show that 
the typical timescale for minihalo photoevaporation is 
$t_{\rm ev} \sim 10-100$ Myr. This is comparable to the recombination
time $t_{\rm rec} = 1/(c_l \alpha_B \langle n_H \rangle) \sim 100$ Myr 
at $z \sim 10$ for a clumpy IGM with $c_l = 10$. This suggests that Jeans smoothing
of the IGM occurs before recombinations have had time to significantly disturb
the reionization process. The amount
by which recombinations occurring within minihalos could delay reionization was studied
by \citet{ciardi/etal:2006} who performed numerical simulations using the results of
\citet{2005MNRAS.361..405I} as a subgrid model for minihalo absorption. For the extreme
case where Jeans smoothing fails to suppress minihalo formation they place $\Delta z \sim 2$
as the upper limit to the redshift delay of reionization induced by minihalos. For the opposite and
more physically realistic case where minihalo formation is heavily suppressed by photoionization
they find only a modest impact on reionization with a volume averaged ionized fraction that is 
$\lesssim 15\%$ lower than the case where minihalo recombinations are ignored.

However, the photoevaporative
process is in reality likely to be more complex than for the
simplified geometries and source lifetimes considered by
\citet{shapiro/etal:2004}, with 
halos over a range of masses clustered in space and arranged within a
``cosmic web'' of filamentary structure. 
Filamentary infall from nearby neutral
gas could replenish halos as they are being evaporated, considerably
extending the photoevaporation process, while ionization from highly
luminous but intermittent starburst galaxies could result in large
clumping factors and stalled minihalo evaporation, considerably
increasing photon consumption and leading to a much more complex
morphology of early H II regions \citep[e.g.,][]{wise/etal:2012} than
is typically envisioned.  

A possible scenario that we have not considered here is the suppression of
gas clumping at early times due to the 
presence of high-redshift X-ray sources \citep[see, e.g.,][]{haiman:2011}.
These may be associated with traditional X-ray sources like supernova or by 
more exotic sources like microquasars \citep{mirabel/etal:2011}. 
X-rays have a small absorption cross-section meaning that a high-redshift  
distribution of X-ray sources would expose the IGM to a nearly uniform source of 
heating, inhibiting minihalo formation and growth at early times \citep{oh/haiman:2003}.
The result is a warm ($T \sim 1000$ K) and weakly ionized IGM that would later become reionized
by the patchy network of star-forming galaxies with softer radiation spectra. 
In this case the clumping factor would already be reduced at the onset of reionization
and the resolution requirements presented here would become less strict. 
Our convergence criteria may therefore be considered the conservative case
where the IGM has not been smoothed by heating processes prior to reionization.

We find that convergence is reached at a dark matter particle mass of $m_{\rm
  dm} \lesssim 50~ M_\odot$. A box size of $L \gtrsim 1$ Mpc is
necessary to sample the IGM for the purpose of modelling absorptions
by small-scale structure. The clumping factors we find from our
converged results are somewhat smaller than the values $c_l\sim
30$ found in early attempts to characterize the clumpiness of the IGM
which did not accurately separate ionized and neutral gas
\citep[e.g.,][]{gnedin/ostriker:1997}, but are higher than the
clumping factors found by \citet{pawlik/etal:2009} at $z\sim 9$, just
before the IGM in those simulations was heated by ionizing
radiation. We attribute this difference to the increased mass
resolution of our simulations, which resolve halo masses down to the
Jeans mass in an {\em unheated} IGM ($\sim 10^4 M_\odot$), as opposed
to that corresponding to the Jeans mass for a photoionized gas
temperature of $\sim 10^4$~K ($\sim 10^8 M_\odot$).  As
pointed out by \citet{pawlik/etal:2009}, the clumping factors they
find at $z\sim 6$, for a patch of the IGM which was ionized
significantly earlier, at $z\sim 9$, 
are converged with respect to the Jeans scale of the heated IGM.
Their value likely approaches the {\em correct}\footnote{The correctness of 
the clumping factor depends on the specific physicalal processes affecting
the evolution of baryons within the IGM and on the particular context to 
which the clumping factor is being used to describe. Here
we use the term correct to refer to the value of the clumping factor that
would be obtained if the simulation in question had infinite resolution.}
value for the 
post-reionization IGM at $z = 6$ because
a long enough time had passed since the gas was ionized for
photoheating to evaporate existing small-scale structure and suppress
accretion onto newly formed dark matter minihalos with masses below
$\sim 10^{8-9}~ M_\odot$, which were resolved in their highest
resolution simulations by $\sim 100-1000$ dark matter particles. 

\begin{figure}
\smjustify
\includegraphics[width=\smwidth]{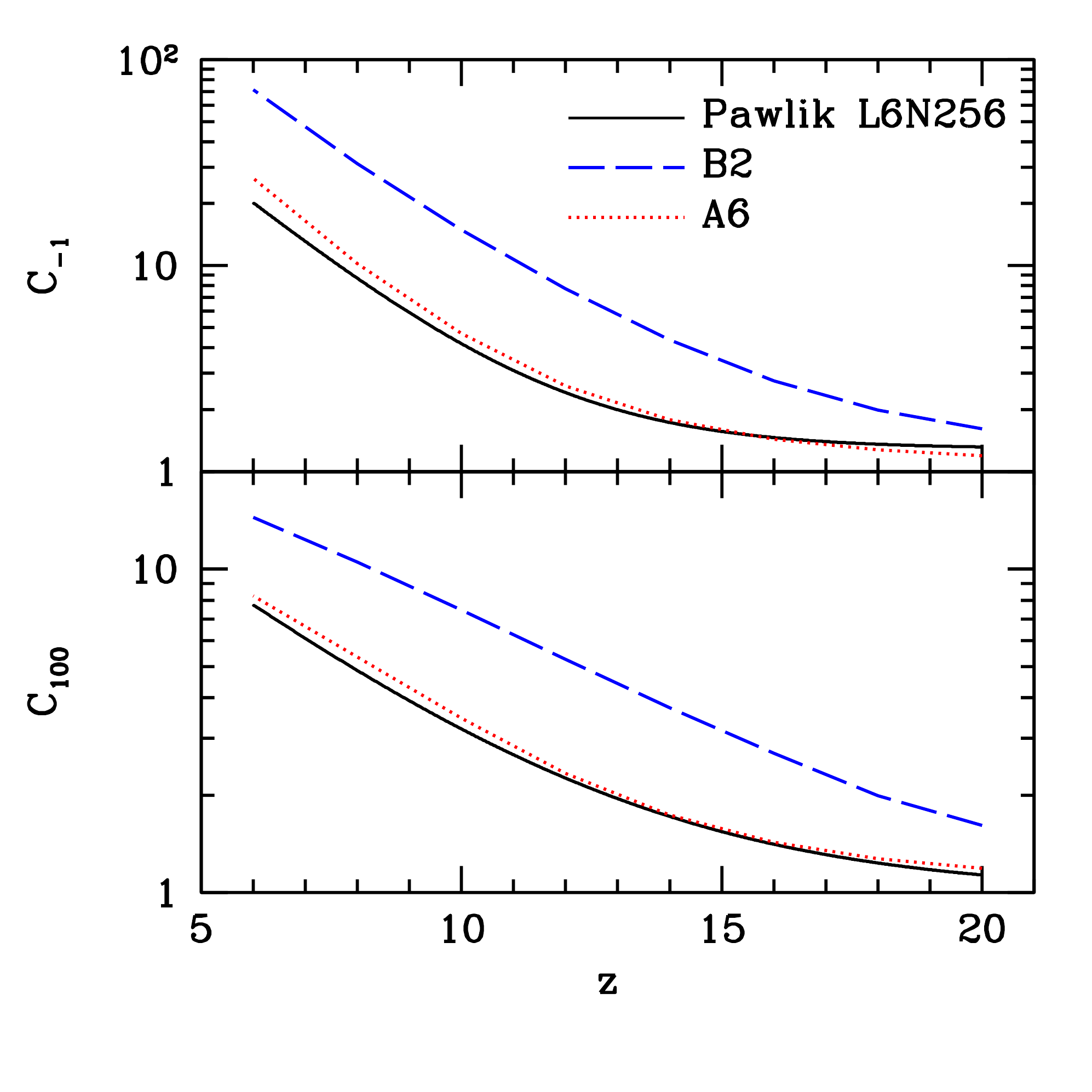}
\vspace{-0.7cm}
\caption{Here we compare clumping factors $C_{-1}$ and $C_{100}$ from our fiducial
  simulation B2 ($512^3$, 0.5 Mpc) to the \citet{pawlik/etal:2009} simulation
  L6N256 ($256^3$, 6 Mpc). The Pawlik et al. curve corresponds to their reference
  simulation that did not include photoionization and has similar
  attributes to our simulation A6 ($256^3$, 8 Mpc). As expected, A6 and L6N256 are in
  good agreement with each other while the higher-resolved simulation
  B2 shows clumping factors a few times larger than L6N256.} 
\label{figure:pawlik}
\vspace{0.2cm}
\end{figure}

\citet{pawlik/etal:2009} demonstrate convergence of their clumping
factor for the heated IGM while noting that convergence would be more difficult to
obtain for an unheated IGM. We have explicitly demonstrated this latter point
here and accordingly show that their values for the clumping factor are likely underestimates
during the initial stages of reionization, by about a factor of a few.
This is illustrated in Figure \ref{figure:pawlik} where we compare clumping factors
from two of our simulations to the unheated simulation L6N256 of Pawlik et al.
Here we plot the clumping factors $C_{-1}$ and $C_{100}$ with this notation being
used to emphasize that these clumping factors are evaluated using all gas below
some density cutoff.  For $C_{-1}$ a physical
density threshold of $n_{\rm crit} = 0.1~{\rm cm}^{-3}$ is used while an 
overdensity threshold of $\Delta_{\rm crit} = 100$ is used for $C_{100}$. These
definitions differ from the definition of $c_l$ in equation (\ref{eq:cl}) that we have been using so far which
involves an average over all ionized gas in the box. In any event, the clumping factors
we find in our fiducial simulation B2 are at all times larger, by a factor of
1.2 at $z = 20$ and 3.5 at $z = 6$ for $C_{-1}$. 
As described above, the clumping factors we obtain at low
redshift are overestimates, and in reality should be closer to $C_{-1}
\sim 6$ or $C_{100} \sim 3$ found by Pawlik et al. in the presence of a photoevaporative
background. Combining these two results, the clumping
factor of the the IGM evolves strongly just after a patch of IGM is
ionized. For ionization at $z=9$, the clumping factor drops from $c_l
\sim 20$ at $z=9$ to a few at $z=6$, depending on the intensity of the
ionizing background -- with larger intensities leading to higher
clumping factors and larger mean free paths. This strong suppression of
the clumping factor due to photoheating was demonstrated by Pawlik et al.
who referred to it as a positive feedback on reionization since it reduces the 
total number of recombinations occurring within small-scale absorption systems.

The results presented here for the inhomogeneity of electron density
in the presence of an ionizing background should serve as a foundation
for more detailed study of radiative transfer and hydrodynamical
effects in the initial 
stages of reionization, including the effects of the initial relative
velocity between baryons and dark matter
\citep[e.g.,][]{tseliakhovich/hirata:2010}, preheating by long mean free path
X-ray photons
\citep[e.g.,][]{ricotti/ostriker:2004,ricotti/etal:2005}, and
photoevaporation \citep[e.g.,][]{shapiro/etal:2004,
  abel/etal:2007}. In simulating all of these processes, it will be
necessary to resolve small-scale structure in the way outlined here. 

\acknowledgments{We offer thanks to P.~R.~Shapiro for helpful discussions
  and to the referee, A.~H.~Pawlik, for his thorough report that 
  improved many aspects of our manuscript. JDE gratefully acknowledges the support of the 
  National Science and Engineering Research Council of Canada.
  The simulations used in this paper were performed on the GPC
  supercomputer at the SciNet HPC Consortium.  SciNet is funded by:
  the Canada Foundation for Innovation under the auspices of Compute
  Canada; the Government of Ontario; Ontario Research Fund - Research
  Excellence; and the University of Toronto.
  }


\begin{thebibliography}{58}
\expandafter\ifx\csname natexlab\endcsname\relax\def\natexlab#1{#1}\fi
\expandafter\ifx\csname href\endcsname\relax
  \def\href#1#2{}\fi
\expandafter\ifx\csname urllinklabel\endcsname\relax
  \def\urllinklabel{[LINK]}\fi
\expandafter\ifx\csname adsurllinklabel\endcsname\relax
  \def\adsurllinklabel{[ADS]}\fi

\bibitem[{{Abel} {et~al.}(2002){Abel}, {Bryan}, \&
  {Norman}}]{2002Sci...295...93A}
{Abel}, T., {Bryan}, G.~L., \& {Norman}, M.~L. 2002, Science, 295, 93


\bibitem[{{Abel} {et~al.}(2007){Abel}, {Wise}, \& {Bryan}}]{abel/etal:2007}
{Abel}, T., {Wise}, J.~H., \& {Bryan}, G.~L. 2007, \apjl, 659, L87


\bibitem[{{Alvarez} \& {Abel}(2012)}]{alvarez/abel:2012}
{Alvarez}, M.~A. \& {Abel}, T. 2012, \apj, 747, 126


\bibitem[{{Alvarez} {et~al.}(2006){Alvarez}, {Bromm}, \&
  {Shapiro}}]{alvarez/etal:2006}
{Alvarez}, M.~A., {Bromm}, V., \& {Shapiro}, P.~R. 2006, \apj, 639, 621


\bibitem[{{Alvarez} {et~al.}(2012){Alvarez}, {Finlator}, \&
  {Trenti}}]{alvarez/etal:2012}
{Alvarez}, M.~A., {Finlator}, K., \& {Trenti}, M. 2012, \apjl, 759, L38


\bibitem[{{Barkana} \& {Loeb}(1999)}]{barkana/loeb:1999}
{Barkana}, R. \& {Loeb}, A. 1999, \apj, 523, 54


\bibitem[{{Barkana} \& {Loeb}(2004)}]{barkana/loeb:2004}
---. 2004, \apj, 609, 474


\bibitem[{{Bolton} \& {Haehnelt}(2007)}]{bolton/haehnelt:2007a}
{Bolton}, J.~S. \& {Haehnelt}, M.~G. 2007, \mnras, 382, 325


\bibitem[{{Bouwens} {et~al.}(2010){Bouwens}, {Illingworth}, {Oesch},
  {Stiavelli}, {van Dokkum}, {Trenti}, {Magee}, {Labb{\'e}}, {Franx},
  {Carollo}, \& {Gonzalez}}]{bouwens/etal:2010}
{Bouwens}, R.~J., {Illingworth}, G.~D., {Oesch}, P.~A., {Stiavelli}, M., {van
  Dokkum}, P., {Trenti}, M., {Magee}, D., {Labb{\'e}}, I., {Franx}, M.,
  {Carollo}, C.~M., \& {Gonzalez}, V. 2010, \apjl, 709, L133


\bibitem[{{Bromm} {et~al.}(2002){Bromm}, {Coppi}, \&
  {Larson}}]{2002ApJ...564...23B}
{Bromm}, V., {Coppi}, P.~S., \& {Larson}, R.~B. 2002, \apj, 564, 23


\bibitem[{{Chiu} {et~al.}(2003){Chiu}, {Fan}, \& {Ostriker}}]{chiu/etal:2006}
{Chiu}, W.~A., {Fan}, X., \& {Ostriker}, J.~P. 2003, \apj, 599, 759


\bibitem[{{Choudhury} {et~al.}(2009){Choudhury}, {Haehnelt}, \&
  {Regan}}]{choudhury/etal:2009}
{Choudhury}, T.~R., {Haehnelt}, M.~G., \& {Regan}, J. 2009, \mnras, 394, 960


\bibitem[{{Ciardi} {et~al.}(2006){Ciardi}, {Scannapieco}, {Stoehr}, {Ferrara},
  {Iliev}, \& {Shapiro}}]{ciardi/etal:2006}
{Ciardi}, B., {Scannapieco}, E., {Stoehr}, F., {Ferrara}, A., {Iliev}, I.~T.,
  \& {Shapiro}, P.~R. 2006, \mnras, 366, 689


\bibitem[{{Couchman} \& {Rees}(1986)}]{couchman/rees:1986}
{Couchman}, H.~M.~P. \& {Rees}, M.~J. 1986, \mnras, 221, 53


\bibitem[{{Fan} {et~al.}(2006){Fan}, {Strauss}, {Becker}, {White}, {Gunn},
  {Knapp}, {Richards}, {Schneider}, {Brinkmann}, \& {Fukugita}}]{Fan06}
{Fan}, X., {Strauss}, M.~A., {Becker}, R.~H., {White}, R.~L., {Gunn}, J.~E.,
  {Knapp}, G.~R., {Richards}, G.~T., {Schneider}, D.~P., {Brinkmann}, J., \&
  {Fukugita}, M. 2006, \aj, 132, 117


\bibitem[{{Finlator} {et~al.}(2012){Finlator}, {Oh}, {{\"O}zel}, \&
  {Dav{\'e}}}]{finlator/etal:2012}
{Finlator}, K., {Oh}, S.~P., {{\"O}zel}, F., \& {Dav{\'e}}, R. 2012, ArXiv
  e-prints


\bibitem[{{Furlanetto} {et~al.}(2004){Furlanetto}, {Zaldarriaga}, \&
  {Hernquist}}]{furlanetto/etal:2004}
{Furlanetto}, S.~R., {Zaldarriaga}, M., \& {Hernquist}, L. 2004, \apj, 613, 1


\bibitem[{{Gnedin} \& {Fan}(2006)}]{gnedin/fan:2006}
{Gnedin}, N.~Y. \& {Fan}, X. 2006, \apj, 648, 1


\bibitem[{{Gnedin} \& {Hui}(1998)}]{gnedin/hui:1998}
{Gnedin}, N.~Y. \& {Hui}, L. 1998, \mnras, 296, 44


\bibitem[{{Gnedin} \& {Ostriker}(1997)}]{gnedin/ostriker:1997}
{Gnedin}, N.~Y. \& {Ostriker}, J.~P. 1997, \apj, 486, 581


\bibitem[{{Greif} {et~al.}(2008){Greif}, {Johnson}, {Klessen}, \&
  {Bromm}}]{greif/etal:2008}
{Greif}, T.~H., {Johnson}, J.~L., {Klessen}, R.~S., \& {Bromm}, V. 2008,
  \mnras, 387, 1021


\bibitem[{{Haardt} \& {Madau}(1996)}]{haardt/madau:1996}
{Haardt}, F. \& {Madau}, P. 1996, \apj, 461, 20


\bibitem[{{Haiman}(2011)}]{haiman:2011}
{Haiman}, Z. 2011, \nat, 472, 47


\bibitem[{{Haiman} {et~al.}(2001){Haiman}, {Abel}, \&
  {Madau}}]{haiman/etal:2001}
{Haiman}, Z., {Abel}, T., \& {Madau}, P. 2001, \apj, 551, 599


\bibitem[{{Haiman} {et~al.}(2000){Haiman}, {Abel}, \&
  {Rees}}]{haiman/etal:2000}
{Haiman}, Z., {Abel}, T., \& {Rees}, M.~J. 2000, \apj, 534, 11


\bibitem[{{Iliev} {et~al.}(2006){Iliev}, {Mellema}, {Pen}, {Merz}, {Shapiro},
  \& {Alvarez}}]{iliev/etal:2006}
{Iliev}, I.~T., {Mellema}, G., {Pen}, U.-L., {Merz}, H., {Shapiro}, P.~R., \&
  {Alvarez}, M.~A. 2006, \mnras, 369, 1625


\bibitem[{{Iliev} {et~al.}(2005{\natexlab{a}}){Iliev}, {Scannapieco}, \& {
  }}]{iliev/etal:2005}
{Iliev}, I.~T., {Scannapieco}, E., \& { }, P.~R. 2005{\natexlab{a}}, \apj, 624,
  491


\bibitem[{{Iliev} {et~al.}(2005{\natexlab{b}}){Iliev}, {Shapiro}, \&
  {Raga}}]{2005MNRAS.361..405I}
{Iliev}, I.~T., {Shapiro}, P.~R., \& {Raga}, A.~C. 2005{\natexlab{b}}, \mnras,
  361, 405


\bibitem[{{Johnson} {et~al.}(2007){Johnson}, {Greif}, \&
  {Bromm}}]{johnson/etal:2007}
{Johnson}, J.~L., {Greif}, T.~H., \& {Bromm}, V. 2007, \apj, 665, 85


\bibitem[{{Komatsu} {et~al.}(2011){Komatsu}, {Smith}, {Dunkley}, {Bennett},
  {Gold}, {Hinshaw}, {Jarosik}, {Larson}, {Nolta}, {Page}, {Spergel},
  {Halpern}, {Hill}, {Kogut}, {Limon}, {Meyer}, {Odegard}, {Tucker}, {Weiland},
  {Wollack}, \& {Wright}}]{komatsu/etal:2011}
{Komatsu}, E., {Smith}, K.~M., {Dunkley}, J., {Bennett}, C.~L., {Gold}, B.,
  {Hinshaw}, G., {Jarosik}, N., {Larson}, D., {Nolta}, M.~R., {Page}, L.,
  {Spergel}, D.~N., {Halpern}, M., {Hill}, R.~S., {Kogut}, A., {Limon}, M.,
  {Meyer}, S.~S., {Odegard}, N., {Tucker}, G.~S., {Weiland}, J.~L., {Wollack},
  E., \& {Wright}, E.~L. 2011, \apjs, 192, 18


\bibitem[{{Kuhlen} \& {Faucher-Gigu{\`e}re}(2012)}]{kuhlen:2012}
{Kuhlen}, M. \& {Faucher-Gigu{\`e}re}, C.-A. 2012, \mnras, 423, 862


\bibitem[{{Madau} {et~al.}(1999){Madau}, {Haardt}, \& {Rees}}]{madau/etal:1999}
{Madau}, P., {Haardt}, F., \& {Rees}, M.~J. 1999, \apj, 514, 648


\bibitem[{{McQuinn} {et~al.}(2011){McQuinn}, {Oh}, \&
  {Faucher-Gigu{\`e}re}}]{mcquinn/etal:2011}
{McQuinn}, M., {Oh}, S.~P., \& {Faucher-Gigu{\`e}re}, C.-A. 2011, \apj, 743, 82


\bibitem[{{Mirabel} {et~al.}(2011){Mirabel}, {Dijkstra}, {Laurent}, {Loeb}, \&
  {Pritchard}}]{mirabel/etal:2011}
{Mirabel}, I.~F., {Dijkstra}, M., {Laurent}, P., {Loeb}, A., \& {Pritchard},
  J.~R. 2011, \aap, 528, A149


\bibitem[{{Miralda-Escud{\' e}} {et~al.}(2000){Miralda-Escud{\' e}},
  {Haehnelt}, \& {Rees}}]{miralda-escude/etal:2000}
{Miralda-Escud{\' e}}, J., {Haehnelt}, M., \& {Rees}, M.~J. 2000, \apj, 530, 1


\bibitem[{{Oesch} {et~al.}(2012){Oesch}, {Bouwens}, {Illingworth}, {Labb{\'e}},
  {Trenti}, {Gonzalez}, {Carollo}, {Franx}, {van Dokkum}, \&
  {Magee}}]{oesch/etal:2012}
{Oesch}, P.~A., {Bouwens}, R.~J., {Illingworth}, G.~D., {Labb{\'e}}, I.,
  {Trenti}, M., {Gonzalez}, V., {Carollo}, C.~M., {Franx}, M., {van Dokkum},
  P.~G., \& {Magee}, D. 2012, \apj, 745, 110


\bibitem[{{Oh} \& {Haiman}(2003)}]{oh/haiman:2003}
{Oh}, S.~P. \& {Haiman}, Z. 2003, \mnras, 346, 456


\bibitem[{{Pawlik} {et~al.}(2009){Pawlik}, {Schaye}, \& {van
  Scherpenzeel}}]{pawlik/etal:2009}
{Pawlik}, A.~H., {Schaye}, J., \& {van Scherpenzeel}, E. 2009, \mnras, 394,
  1812


\bibitem[{{Peebles} \& {Dicke}(1968)}]{peebles/dicke:1968}
{Peebles}, P.~J.~E. \& {Dicke}, R.~H. 1968, \apj, 154, 891


\bibitem[{{Prochaska} {et~al.}(2009){Prochaska}, {O'Meara}, \&
  {Worseck}}]{prochaska/etal:2009}
{Prochaska}, J.~X., {O'Meara}, J.~M., \& {Worseck}, G. 2009, arXiv:0912.0292


\bibitem[{{Reed} {et~al.}(2007){Reed}, {Bower}, {Frenk}, {Jenkins}, \&
  {Theuns}}]{reed/etal:2007}
{Reed}, D.~S., {Bower}, R., {Frenk}, C.~S., {Jenkins}, A., \& {Theuns}, T.
  2007, \mnras, 374, 2


\bibitem[{{Ricotti} \& {Ostriker}(2004)}]{ricotti/ostriker:2004}
{Ricotti}, M. \& {Ostriker}, J.~P. 2004, \mnras, 352, 547


\bibitem[{{Ricotti} {et~al.}(2005){Ricotti}, {Ostriker}, \&
  {Gnedin}}]{ricotti/etal:2005}
{Ricotti}, M., {Ostriker}, J.~P., \& {Gnedin}, N.~Y. 2005, \mnras, 357, 207


\bibitem[{{Shapiro} \& {Giroux}(1987)}]{shapiro/giroux:1987}
{Shapiro}, P.~R. \& {Giroux}, M.~L. 1987, \apj, 321, L107


\bibitem[{{Shapiro} {et~al.}(1994){Shapiro}, {Giroux}, \&
  {Babul}}]{shapiro/etal:1994}
{Shapiro}, P.~R., {Giroux}, M.~L., \& {Babul}, A. 1994, \apj, 427, 25


\bibitem[{{Shapiro} {et~al.}(2004){Shapiro}, {Iliev}, \&
  {Raga}}]{shapiro/etal:2004}
{Shapiro}, P.~R., {Iliev}, I.~T., \& {Raga}, A.~C. 2004, \mnras, 348, 753


\bibitem[{{Shull} {et~al.}(2012){Shull}, {Harness}, {Trenti}, \&
  {Smith}}]{shull/etal:2012}
{Shull}, J.~M., {Harness}, A., {Trenti}, M., \& {Smith}, B.~D. 2012, \apj, 747,
  100


\bibitem[{{Springel}(2005)}]{springel/etal:2005}
{Springel}, V. 2005, \mnras, 364, 1105


\bibitem[{{Storrie-Lombardi} {et~al.}(1994){Storrie-Lombardi}, {McMahon},
  {Irwin}, \& {Hazard}}]{storrie-lombardi/etal:1994}
{Storrie-Lombardi}, L.~J., {McMahon}, R.~G., {Irwin}, M.~J., \& {Hazard}, C.
  1994, \apjl, 427, L13


\bibitem[{{Tegmark} {et~al.}(1997){Tegmark}, {Silk}, {Rees}, {Blanchard},
  {Abel}, \& {Palla}}]{tegmark/etal:1997}
{Tegmark}, M., {Silk}, J., {Rees}, M.~J., {Blanchard}, A., {Abel}, T., \&
  {Palla}, F. 1997, \apj, 474, 1


\bibitem[{{Trenti} {et~al.}(2010){Trenti}, {Stiavelli}, {Bouwens}, {Oesch},
  {Shull}, {Illingworth}, {Bradley}, \& {Carollo}}]{trenti/etal:2010}
{Trenti}, M., {Stiavelli}, M., {Bouwens}, R.~J., {Oesch}, P., {Shull}, J.~M.,
  {Illingworth}, G.~D., {Bradley}, L.~D., \& {Carollo}, C.~M. 2010, \apjl, 714,
  L202


\bibitem[{{Tseliakhovich} \& {Hirata}(2010)}]{tseliakhovich/hirata:2010}
{Tseliakhovich}, D. \& {Hirata}, C. 2010, \prd, 82, 083520


\bibitem[{{Warren} {et~al.}(2006){Warren}, {Abazajian}, {Holz}, \&
  {Teodoro}}]{warren/etal:2006}
{Warren}, M.~S., {Abazajian}, K., {Holz}, D.~E., \& {Teodoro}, L. 2006, \apj,
  646, 881


\bibitem[{{Wise} \& {Abel}(2008)}]{wise/abel:2008}
{Wise}, J.~H. \& {Abel}, T. 2008, \apj, 685, 40


\bibitem[{{Wise} \& {Cen}(2009)}]{wise/cen:2009}
{Wise}, J.~H. \& {Cen}, R. 2009, \apj, 693, 984


\bibitem[{{Wise} {et~al.}(2012){Wise}, {Turk}, {Norman}, \&
  {Abel}}]{wise/etal:2012}
{Wise}, J.~H., {Turk}, M.~J., {Norman}, M.~L., \& {Abel}, T. 2012, \apj, 745,
  50


\bibitem[{{Yoshida} {et~al.}(2003){Yoshida}, {Abel}, {Hernquist}, \&
  {Sugiyama}}]{2003ApJ...592..645Y}
{Yoshida}, N., {Abel}, T., {Hernquist}, L., \& {Sugiyama}, N. 2003, \apj, 592,
  645


\bibitem[{{Yoshida} {et~al.}(2007){Yoshida}, {Oh}, {Kitayama}, \&
  {Hernquist}}]{yoshida/etal:2007}
{Yoshida}, N., {Oh}, S.~P., {Kitayama}, T., \& {Hernquist}, L. 2007, \apj, 663,
  687


\end{thebibliography}
\end{document}